\documentclass[aps, pra, twocolumn, superscriptaddress]{revtex4-1}
\usepackage{amssymb, amsthm}
\usepackage{amsmath}
\usepackage{color}
\usepackage{graphicx}
\usepackage{hyperref}
\usepackage{comment}

\newcommand{\ket}[1]{| #1 \rangle}
\newcommand{\bra}[1]{\langle #1 |}
\newcommand{\Var}{\mathrm{Var}}

\newcommand{\mytitle}{One-dimensional spin-1/2 fermionic gases with two-body losses: weak dissipation and spin conservation}

\begin{document}
  
\title{\mytitle}      
   
\author{Lorenzo Rosso}
\affiliation{Universit\'e Paris-Saclay, CNRS, LPTMS, 91405 Orsay, France}

\author{Davide Rossini}
\affiliation{Dipartimento di Fisica dell'Universit\`a di Pisa and INFN, Largo Pontecorvo 3, 56127 Pisa, Italy}

\author{Alberto Biella}
\affiliation{Universit\'e Paris-Saclay, CNRS, LPTMS, 91405 Orsay, France}
\affiliation{INO-CNR BEC Center and Dipartimento di Fisica, Universit\`a di Trento, 38123 Povo, Italy}

\author{Leonardo Mazza}
\email{leonardo.mazza@universite-paris-saclay.fr}
\affiliation{Universit\'e Paris-Saclay, CNRS, LPTMS, 91405 Orsay, France}

\begin{abstract}
  We present a theoretical analysis of the dynamics of a one-dimensional spin-1/2 fermionic gas subject to weak two-body losses. 
  Our approach highlights the crucial role played by spin conservation in the determination of the full time evolution. 
  We focus in particular on the dynamics of a gas that is initially prepared in a Dicke state with fully-symmetric spin wavefunction, in a band insulator, or in a Mott insulator.
  In the latter case, we investigate the emergence of a steady symmetry-resolved purification of the gas.
  Our results could help the modelisation and understanding of recent experiments with alkaline-earth(-like) gases like ytterbium or fermionic molecules.
\end{abstract}

\maketitle

\section{Introduction}

Experiments with ultra-cold gases are often regarded as paradigmatic studies of closed many-body quantum systems~\cite{LangenRev_2015}; 
yet, they always suffer from the continuous leakage of particles into the vacuum chamber.
In general, losses are responsible for decoherence and for the disappearance of quantum coherence~\cite{Zurek_2003}. 
Several studies have pointed out that they can also induce interesting effects: they can be used as a diagnostic tool for strong correlations~\cite{Tolra_2004, Kraemer_2006, Pollack_2009, Baur_2010}, 
they can purify and cool the gas~\cite{Grisins_2016, Bouchoule_2018, Schemmer_2018, Dogra_2019}, or  induce strong quantum correlations~\cite{Syassen_2008, GarciaRipoll_2009, Kantian_2009, Daley_2009,  Roncaglia_2010}.  
These effects are just a simple instance of
the fact that in most situations the coupling to an environment, if properly engineered, can be beneficial and can be exploited for quantum technology purposes~\cite{Almut_2000, Diehl_2008, Verstraete_2009, Diehl_2011, Iemini_2016}.
The correct theoretical modelisation of the quantum dynamics induced by losses has thus emerged as an important problem and has recently attracted widespread attention~\cite{Kordas_2015, Goto_2020, Bouchoule_2020b, Rossini_2020, Bouchoule_2021, Nakagawa_2021}.
 
Fermionic gases trapped in one-dimensional systems and subject to two-body losses feature a class of highly-entangled stationary states with a fully symmetric Dicke-like spin wavefunction, that could have important scientific and technological applications~\cite{FossFeig_2012}.
They have been the object of several experiments with molecular~\cite{Yan_2013, Zhu_2014} and atomic gases~\cite{Sponselee_2018}, which however have not been able to certify the properties of the realised stationary state. 
Spin is conserved during the dissipative dynamics, and this is crucial for determining the steady properties. 
Various theoretical articles have addressed several aspects of the model and of its dynamics~\cite{Nakagawa_2020, Nakagawa_2021} but the impact of spin conservation on the full dynamics, beyond determining its stationary properties, has not been understood yet.

In this article we present a simple theoretical framework for describing the lossy dynamics of a one-dimensional fermionic gas with two-body losses that takes into account the exact conservation of spin.
Even if most of the attention so far has focused on the strongly-dissipative regime, we address here the weakly-dissipative limit, which does not spoil the appearance of the highly-entangled stationary states.
We highlight the crucial role played by spin in causing a non-trivial relaxation dynamics that affects in a qualitative way the long-time behaviour. The simplicity of the approach is a first step towards the modelisation of experimental data, which could be obtained using alkaline-earth (e.g.~strontium) or alkaline-earth like (e.g.~ytterbium~\cite{Sponselee_2018}) atoms in the excited metastable state. 
We expect that our results may open the path to the realisation of a consistent theory for the fermionic dynamics in the Zeno limit, as recently done for bosons~\cite{Rossini_2020}. 

The article is organised as follows. In Sec.~\ref{Sec:setup} we describe the setup. In Sec.~\ref{Sec:Pop_dynamics} we derive a dynamical equation for the evolution of the particle density. Next, we compare our theoretical predictions with numerical simulations for three different classes of initial states: Dicke states with fully-symmetric spin wavefunction (Sec.~\ref{Sec:Dicke}), band insulator (Sec.~\ref{Sec:Band}) and Mott insulators (Sec.~\ref{Sec:Mott}). In Sec.~\ref{Sec:Interactions} we discuss the effect of weak interactions. Finally, in Sec~\ref{Sec:Concl} we draw our conclusions.

\section{The setup}
\label{Sec:setup}

We consider a gas of spin-1/2 fermions trapped in a one-dimensional optical lattice with two-body contact interaction and two-body on-site losses.
We introduce the fermionic operators $\hat c_{j,\sigma}^{(\dagger)}$, which satisfy canonical anticommutation relations, and the Hamiltonian of the Hubbard model, which describes the gas in the single-band approximation:
\begin{equation}
  \hat H = - J \sum_{j} \sum_{ \sigma = \uparrow, \downarrow} \left( \hat c_{j, \sigma}^\dagger \, \hat c_{j+1, \sigma}+ {\rm H.c.} \right) + U \sum_j \hat n_{j, \uparrow} \, \hat n_{j, \downarrow}.
\end{equation}
Here, $J$ is the hopping amplitude, $U$ is the interaction strength, and $\hat n_{j,\sigma} = \hat c^\dagger_{j,\sigma} \, \hat c_{j, \sigma}$ is the spin-resolved on-site lattice density operator. 

The presence of local two-body losses is accounted for by the jump operators $\hat L_j = \sqrt{\gamma} \,\hat c_{j, \uparrow} \hat c_{j, \downarrow} $, where $\gamma$ is the loss rate. The dynamics of the full density matrix $\rho(t)$ is described by a Lindblad master equation:
\begin{equation}
  \dot \rho(t) = - \frac{i}{\hbar} \left[ \hat H, \rho(t) \right] + \sum_j \hat L_j \rho(t) \hat L_j^\dagger - \frac 12 \left\{ \hat L_j^\dagger \hat L_j, \rho(t) \right\},
  \label{Eq:MEQ}
\end{equation}
where $[ \cdot, \cdot ]$ denotes the commutator and $\{ \cdot, \cdot \}$ the anticommutator.
In the experimental situations that we want to model~\cite{Sponselee_2018}, i.e.~those where losses are intrinsic, the ratio $\gamma/U$ is determined by atomic (or molecular) properties, and is of the order of the unity; the ratio $\gamma/J$ is instead tunable at will by modulating the strength of the optical lattice potential. We note that there are experiments in which the ratio $\gamma/U$ can be tuned at will by means, for instance, of laser light, as it has been done in Ref.~\cite{Tomita_2017}.

We introduce the operator associated with the total spin of the gas: $\hat{\vec S} = \frac{\hbar}{2} \sum_{j, \sigma, \sigma'} \hat c^\dagger_{j,\sigma} \, \vec \sigma_{\sigma \sigma'} \, \hat c_{j, \sigma'}$, where $\vec \sigma_{\sigma \sigma'}$ is a vector whose components are the Pauli matrices. Since any two-body loss does not change the spin of the gas along any direction (the doubly occupied state has spin $0$),
the dynamics expressed by Eq.~\eqref{Eq:MEQ} is spin conserving and the expectation value of the spin along any direction, $\vec S \cdot \vec n$, is a constant of motion. 
It follows that also $\langle \hat S^2 \rangle$ is a constant of motion. 
The main purpose of this article is to show how the presence of this constraint influences the dynamics.

\section{Population dynamics for weak dissipation}
\label{Sec:Pop_dynamics}

We focus on the simplest experimental observable, $\hat N = \sum_{j,\sigma} \hat n_{j,\sigma}$, and characterise how the number of fermions contained in the sample decreases in time because of loss processes. We will use the notation $\langle \hat A \rangle_t$ to denote the time-dependent expectation value of the observable $\hat A$, namely $\langle \hat A \rangle_t \doteqdot \text{tr}\big[ \rho(t) \, \hat A \big]$. 

With simple algebraic passages, it is possible to show that $N(t) \doteqdot \langle \hat N \rangle_t$ obeys the following equation, which has an intuitive physical meaning:
\begin{equation}
  \frac{\rm d}{\mathrm d t}N(t) = - 2 \gamma \,\Big\langle \sum_j \hat n_{j, \uparrow} \, \hat n_{j, \downarrow}\Big\rangle_t.
  \label{Eq:N:Diss:1}
\end{equation}
Since we cannot treat in an exact analytical way the r.h.s.~of this equation, we perform a series of approximations that are well justified in the limit of weak dissipation, $\hbar \gamma \ll J$. Note that this is also the limit of weak interactions, $U \ll J$, and the first approximation that we perform, discussed extensively in Sec.~\ref{Sec:Interactions}, consists in completely neglecting interactions.

Using a semiclassical reasoning, we observe that in the limit $\hbar\gamma \ll J$, losses act on time scales that are much longer than those characterising the unitary time evolution. In between two loss processes, the long unitary dynamics has acted  and averaged out any time-dependent physical operator or correlator.
We thus employ a \textit{time-dependent stationarity condition} and assume that the system is always in a stationary state of the Hamiltonian, and that particle losses are responsible for a dynamics that explores this subspace of the state space.
This theoretical approach follows from several ideas put forward in the context of weakly-dissipative systems~\cite{Lange_2018, Mallaya_2019}, that have been largely employed in recent theoretical studies of lossy systems~\cite{Bouchoule_2020b, Rossini_2020}.

In practice, 
we focus on the operator that is responsible for losses,  
$\langle \sum_j  \hat n_{j, \uparrow} \, \hat n_{j, \downarrow} \rangle_t$, and expand it in the basis of plane waves, $\hat c_{k, \sigma} = L^{- \frac 12} \sum_j e^{-ikj} \hat c_{j, \sigma}$, that are the eigenmodes of the free-fermion Hamiltonian dynamics, obtaining:
\begin{equation}
 \frac 1L
  \sum_{k,q,w,z} \sum_n \delta_{k+q- w-z,2\pi n} \langle \hat c^\dagger_{k, \uparrow} \, \hat c_{w, \uparrow} \, \hat c^\dagger_{q, \downarrow} \, \hat c_{z, \downarrow} \rangle_t \,.
  \label{Eq:Expectation:NN}
\end{equation}
The Kronecker delta ensures that only momentum-conserving correlators (modulus $2 \pi$) are considered. 
The Hamiltonian time-evolution of the correlators in~\eqref{Eq:Expectation:NN} is easily written:
\begin{equation}
  \langle \hat c^\dagger_{k, \uparrow} \hat c_{w, \uparrow} \hat c^\dagger_{q, \downarrow}  \hat c_{z, \downarrow} \rangle_t \! = \! e^{- \frac{i}{\hbar} \left( E_k + E_q - E_w-E_z \right) t}
  \langle \hat c^\dagger_{k, \uparrow} \hat c_{w, \uparrow} \hat c^\dagger_{q, \downarrow}  \hat c_{z, \downarrow} \rangle_0,
\end{equation}
where $E_k=-2J \cos(k)$ is the energy eigenvalue associated with the $k$-th mode of the free-fermion Hamiltonian.
The request that the system explores only stationary states requires us to keep only
the energy-conserving correlators, because their expectation value does not depend on time.

\begin{widetext}
This leads to an expression that can be further simplified by taking into account the conserved quantity $\hat S^2$ (see Appendix~\ref{App:Derivation:Dyn} for the explicit calculations):
\begin{equation}
  \dot N(t)  = - \frac{2 \gamma}{L} \left[
    \frac{N(t)^2}{4}+\frac{N(t)}{2}+ \frac{\text{Var}N_t}{4} - \frac{\langle \hat S^2 \rangle_0}{\hbar^2}  - \langle\hat \Pi \rangle_t
    +\langle \hat \Sigma_{\frac \pi 2} \rangle_t + 
    \langle \hat T_u \rangle_t
    \right] , 
    \label{Eq:Dynamical}
\end{equation}
with 
\begin{subequations}
\begin{gather}
\Var N_t = \langle \hat N^2 \rangle_t- \langle \hat N\rangle_t^2, \qquad \hat \Pi = \sum_k \hat n_{k, \uparrow} \, \hat n_{k, \downarrow}, \qquad
    \hat \Sigma_{\frac \pi2} = \sum_{k \neq q, \; k \neq \pi-q} 
    \hat c_{k, \uparrow}^\dagger 
    \hat c_{q, \uparrow}
    \hat c_{\pi- k, \downarrow}^\dagger 
    \hat c_{\pi-q, \downarrow}, \\
    \hat T _u = \sum_{\delta k \in \left[0, \frac \pi 2 \right]} \left( 
    \hat c_{\frac \pi 2 + \delta k, \uparrow}^\dagger \hat c_{- \frac \pi 2 - \delta k, \uparrow} \hat c_{ \frac \pi 2-\delta k, \downarrow}^\dagger \hat c_{- \frac \pi 2 + \delta k, \downarrow}
    + 
    \hat c_{\frac \pi 2 - \delta k, \uparrow}^\dagger \hat c_{- \frac \pi 2 - \delta k, \uparrow} \hat c_{ \frac \pi 2+\delta k, \downarrow}^\dagger \hat c_{- \frac \pi 2 + \delta k, \downarrow}
    +H.c.\right). 
\end{gather}
\end{subequations}
Eq.~\eqref{Eq:Dynamical} is the main result of our study: it highlights the crucial interplay between the number of particles, its variance, the spin of the gas, and 
various correlators of the gas in momentum space. In particular, $\hat \Pi$ is a density-density correlator, $\hat \Sigma_{\frac \pi 2}$ takes into account correlators that are symmetric with respect to the center of the band, located at $k = \pm \pi/2$ (note that in this operator momenta are defined mod $2 \pi$ to  restrict them to the first Brillouin zone), and $\hat T_u$ considers umklapp terms, where the difference in momenta is equal to $\pm 2 \pi$. The presence of the two latter operators is a lattice effect: the symmetry of the band with respect to $k=\pm \pi/2$ is not present for a quadratic band in the continuum limit, where $E_k \propto k^2$; similarly umklapp processes exist only in discrete systems.
\end{widetext}

\subsection{The thermodynamic limit}

The terms which appear in the r.h.s.~of Eq.~\eqref{Eq:Dynamical} have different scalings in the thermodynamic limit. We divide both the r.h.s.~and l.h.s.~of Eq.~\eqref{Eq:Dynamical} by $L$ and focus on intensive quantities, whose limit is finite in the thermodynamic limit, which we simply indicate as $\lim_{L \to \infty}$. 
We define the lattice density $n(t) =  \lim_{L \to \infty} N(t)/L$, the lattice spin density $s^2_0 =  \lim_{L \to \infty} \langle \hat S^2 \rangle_0 / L^2$ and the correlator $\sigma_{\frac \pi 2}(t) =  \lim_{L \to \infty}\langle \hat \Sigma_{\frac{\pi}{2}}\rangle_t / L^2$. 
It is expected that $\text{Var} N_t$ and $N(t)$ scale to zero once divided by $L^2$; a similar result is expected for $\langle \hat \Pi \rangle$ and $\langle \hat T_u \rangle$ because they are the sum of $L$ terms. We obtain the simpler equation:
\begin{equation}
  \dot n(t) = - 2 \gamma \left[ \frac{n(t)^2}{4} - \frac{s_0^2}{\hbar^2} + \sigma_{\frac \pi 2}(t)\right].
\end{equation}
Note that a finite spin $\langle \hat S^2 \rangle_0 \neq 0$ could have zero value $s_0=0$ in the thermodynamic limit.

We now propose an argument demonstrating that $\sigma_{\frac \pi 2}(t)=0$ using the fact that the local properties of the system in the thermodynamic limit can be discussed also within the framework of the time-dependent generalised Gibbs ensemble (GGE)~\cite{Lange_2018, Rossini_2020}.
This is a stronger approximation with respect to that used at the beginning of this section to derive Eq.~\eqref{Eq:Dynamical} and requires that not only the system is always in a stationary state of the Hamiltonian, but also in that specific class of states that are GGE.
These states describe the local properties of systems in the thermodynamic limit after a long unitary time evolution. In the case of non-interacting fermions, a GGE is a Gaussian state that is factorised in momentum space.
We present a detailed derivation in Appendix~\ref{Sec:Gaussian} and report here the result:
\begin{equation}
  \dot N(t) = -\frac{2 \gamma}{L} \left[ \frac{N(t)^2}{4} - \frac{\langle \hat S_x\rangle^2+\langle \hat S_y\rangle^2+\langle \hat S_z\rangle^2}{\hbar^2}\right].
\end{equation}
If we focus on intensive quantities and address the thermodynamic limit we observe that:
\begin{equation}
  \lim_{L \to \infty} \frac{\langle \hat S^2 \rangle}{L^2} = \lim_{L\to \infty} \frac{\langle \hat S_x\rangle^2+\langle \hat S_y\rangle^2+\langle \hat S_z\rangle^2}{L^2},
\end{equation}
their difference $\sum_i \langle \hat S_i^{2}\rangle-\langle \hat S_i\rangle^2$ being only $\sum_i \Var S_i$, which is subleading.
We obtain the equation:
\begin{equation}
  \dot n(t) = - 2 \gamma \left[ \frac{n(t)^2}{4} - \frac{s_0^2}{\hbar^2} \right].
  \label{Eq:Dynamical:TL}
\end{equation}
By employing this GGE approximation we thus see that $\sigma_{\frac \pi 2}=0$ and that can be neglected. Moreover, this discussion has the advantage to show explicitly that Eq.~\eqref{Eq:Dynamical:TL} does not take into account exactly the spin conservation.

Eq.~\eqref{Eq:Dynamical} is thus a more refined version of Eq.~\eqref{Eq:Dynamical:TL} because it includes finite-size corrections. 
Whereas this might seem an unnecessary overshooting,
testing this theory with numerical tools is demanding and we will present state-of-the-art numerical simulations for lattices up to $L=10$.
For these lattice lengths, in several cases the dynamics predicted by Eq.~\eqref{Eq:Dynamical:TL} is recognisable only at short times. 
This higher accuracy comes at the price of introducing several new variables, 
for which we have not been able to write satisfactory and simple dynamical equations;
when necessary, we will show how to treat them.

\subsection{Continuum limit}

The study presented so far can be easily extended to a setup without the optical lattice of length $L$ (see also Ref.~\cite{FossFeig_2012}). 
Introducing the fermionic fields $\hat \Psi_{\sigma}(x)$, with $\sigma=\uparrow,\downarrow$, the atomic mass $m$ and the interaction parameter $g$,
the Hamiltonian reads:
\begin{eqnarray}
  \hat H_c = \int \sum_\sigma 
  \hat \Psi_\sigma^\dagger(x) \left( - \frac{\hbar^2}{2m} \partial_x^2 \right) \hat \Psi_\sigma(x) \,
       {\rm d}x + \nonumber \\ +g \int \hat \Psi_\uparrow^\dagger(x) \, \hat \Psi_\uparrow(x) \, \hat \Psi_\downarrow^\dagger(x) \, \hat \Psi_\downarrow(x) \, {\rm d}x.
\end{eqnarray}
In order to include loss processes  we introduce the jump operators $\hat J(x) = \sqrt{\xi} \hat \Psi_\uparrow(x) \hat \Psi_\downarrow(x)$, where $\xi$ is the rate associated with two-body losses.  The full Lindblad dynamics reads:
\begin{align}
  \dot \rho(t) = & - \frac{i}{\hbar}
  \big[ \hat H_c, \rho(t) \big] + \nonumber \\ &+\int 
 \hat J(x) \, \rho(t) \, \hat J(x)^\dagger - \frac{1}{2}
 \left\{ \hat J(x)^\dagger \, \hat J(x) , \rho(t) \right\} {\rm d}x.
\end{align}

We now focus on the weakly dissipative limit, characterised by a loss rate that is subleading with respect to the kinetic energy: $\xi  \ll \hbar n/m $, where $n$ is the gas density. Note that this inequality cannot be satisfied at all times for a gas that loses completely its population; however, as we will see, the problem that we are studying is characterised, in certain regimes, by a finite steady density.

With calculations similar to those presented above, one obtains the following dynamical equation for the population $N(t) = \langle\int \sum_\sigma \hat \Psi^\dagger_\sigma(x) \, \hat \Psi_{\sigma}(x) \, {\rm d }x \rangle_t$ in the weakly-dissipative limit: 
\begin{equation}
  \dot N(t) =  - \frac{2 \xi}{L} \left[
  \frac{N(t)^2}{4}+\frac{N(t)}{2}+ \frac{\text{Var}N_t}{4} - \frac{\langle \hat S^2 \rangle_0}{\hbar^2}  - \langle\hat \Pi \rangle_t
  \right].
\end{equation}
The above equation is identical to Eq.~\eqref{Eq:Dynamical}, where $\gamma$ has been replaced with $\xi$. The definitions of the total spin ${\hat S}^2$ and of the $\hat \Pi$ observables are trivial generalisations of those presented above for a lattice.
Operators $\hat \Sigma_{\frac \pi 2}$ and $\hat T_u$ instead do not appear because, as we have already seen, they are a lattice effect.
In the rest of the paper we will focus only on the lattice problem, and several results can be easily generalised to the continuum case.

\section{Dicke states and stationary populations}
\label{Sec:Dicke}

The first test for the population equation~\eqref{Eq:Dynamical} consists in its application to states whose spin part is fully symmetric like in a Dicke state, and which are characterised by a spin quantum number $S=N/2$.
It has been pointed out in Ref.~\cite{FossFeig_2012} that Dicke states (we use from now on this short name to indicate any state with $S = N/2$) are dark states of the dissipative dynamics: since spin is conserved, and the minimal number of fermions that is necessary to create a spin-$s$ state is $2s$, no particle can be lost from a Dicke state without changing the spin quantum number -- that is, the loss cannot take place. 

Eq.~\eqref{Eq:Dynamical} predicts that the population of Dicke states does not evolve in time. 
In order to prove this, we introduce the notation
$\ket{{\rm D}_N}$ for a generic Dicke state with $N$ particles and spin $S=N/2$; the orbital part of the wavefunction can be arbitrary, provided it is fully antisymmetric.
We show in Appendix~\ref{App:Dicke:Relations} that for a generic linear superposition of Dicke states,
$\ket{\Psi_{\rm D}}=\sum_N c_N \ket{{\rm D}_N}$, the following properties hold:
\begin{subequations}
  \label{eq:Dicke}
  \begin{gather}
    \label{eq:Dicke1}
    \bra{\Psi_{\rm D}} \hat S^2 \ket{\Psi_{\rm D}} = \frac{\hbar^2}{2}\left(\frac{N^2+ \text{Var}N}{2}+N\right) ,\\
    \label{eq:Dicke2}
    \bra{\Psi_{\rm D}} \hat \Pi \ket{\Psi_{\rm D}} = 0,
    \; \, 
    \bra{\Psi_{\rm D}} \hat \Sigma_{\frac \pi2} \ket{\Psi_{\rm D}} = 0,
    \; \,
    \bra{\Psi_{\rm D}} \hat T_u \ket{\Psi_{\rm D}} = 0.
  \end{gather}
\end{subequations}
From these properties we can deduce that Dicke states are stationary states of the dynamics: $\dot N(t) = 0$. 

Eq.~\eqref{Eq:Dynamical:TL} takes into account the spin conservation in the thermodynamic limit, and predicts a stationary density 
\begin{equation}
  n_\infty = \frac{2}{\hbar} s_0. 
\end{equation}
Dicke states satisfy this relation; indeed, from the above formulas we obtain: 
\begin{eqnarray}
  \frac{s_0^2}{\hbar^2} &=&
  \lim_{L \to \infty} \frac{\bra{\Psi_{\rm D}} \hat S^2 \ket{\Psi_{\rm D}}}{\hbar^2L^2} \cr
  &&\cr
  &=&  \lim_{L \to \infty}  \left(\frac{N^2 + \Var N+2N}{4L^2}\right)  = \frac{n^2}{4} .
\end{eqnarray}

\section{Dynamics from a band insulator}
\label{Sec:Band}
We now discuss the dissipative dynamics starting from a band insulator, $\ket{\Psi_{\rm BI}} = \prod_{j} \hat c^\dagger_{j, \uparrow} \, \hat c^\dagger_{j, \downarrow} \ket{0}$, where any lattice site is doubly occupied and the initial population is $2L$. The system is in a spin-0 state, $\langle \hat S^2 \rangle = 0$, and a simple calculation shows that $\langle \hat \Pi \rangle_0 = L$, $\langle \hat \Sigma_{\frac \pi2}\rangle_0=0$, and $\langle \hat T_u \rangle_0=0$. The prediction for the dynamics of the lattice density $n(t)$ in the thermodynamic limit is easily obtained from Eq.~\eqref{Eq:Dynamical:TL}:
\begin{equation}
 n(t)= \frac{2} {1+ \gamma t}.
 \label{Eq:n:BI:TL}
\end{equation}
The full solution of Eq.~\eqref{Eq:Dynamical} is more challenging because it is not obvious how to give a prediction for the time-dependence of $\Var N_t$, $\langle \hat \Pi \rangle_t$, $\langle \hat \Sigma_{\frac \pi 2}\rangle_t$, and $\langle \hat T_u \rangle_t$ (we could not derive closed expressions for their time-derivatives and a Gaussian expansion gives wrong predictions, possibly because here we are looking for beyond-Gaussian effects).

\begin{figure}[t]
  \includegraphics[width=\columnwidth]{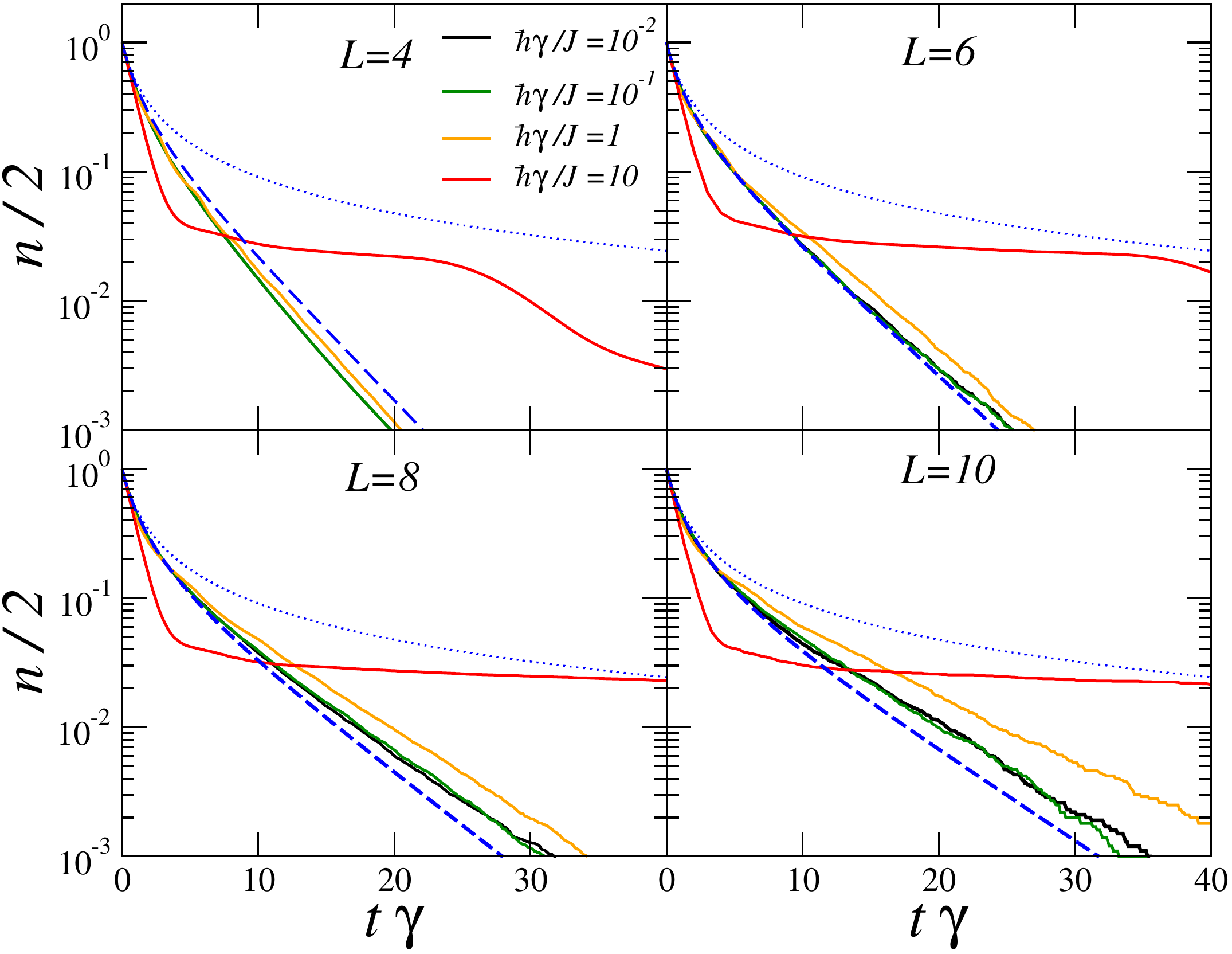}
  \caption{Dissipative dynamics of the normalised density of a band insulator for $L=4$, $6$, $8$ and $10$.  The various colors refer to different dissipation strengths, from $\hbar\gamma/J=10^{-2}$ to $\hbar\gamma/J=10$ (see legend). The thin blue dotted curve is the prediction for the thermodynamic limit in Eq.~\eqref{Eq:n:BI:TL} whereas the thick blue dashed curve is Eq.~\eqref{Eq:BI:n:Theory}.
    The latter faithfully describes the weakly-dissipative limit even at small sizes. 
    The plot highlights the collapse of the curves for $\hbar \gamma/J = 10^{-2}$, $10^{-1}$ and $1$. On the other hand, the appearance of a new behaviour in the strongly-dissipative Zeno limit is evident.}
  \label{Fig:n:BI:LScaling}
\end{figure}

We can use Eq.~\eqref{Eq:Dynamical} to get insights into the long-time dynamics of a finite system, since when $N(t)$ tends to zero we have $N(t)^2 \ll N(t)$. In this limit it is possible to model the number of particles as a Bernoullian distribution, where with probability $p$ the system has 2 particles, and with probability $1-p$ it is empty. For such a distribution, $N = 2p$, $\Var N \sim 4 p$ and thus we estimate that, in the long-time limit, $\Var N_t \sim 2N(t)$. 
Concerning $\langle \hat \Pi \rangle$, we can bound it in the following way: since $\hat \Pi$ is a non-negative operator, $\langle \hat \Pi \rangle \geq 0$; since $\hat n_{k, \uparrow} \, \hat n_{k, \downarrow} \leq \hat n_k/2$, we can write that $\langle \hat \Pi \rangle < N/2$.
If we neglect the contributions from $\hat \Sigma_{\frac \pi2}$ and $\hat T_u$ (which is justified \textit{a posteriori} by numerical simulations), we thus obtain that the long-time scaling is exponential: $N(t) \sim \exp (- t/\tau )$, with $L/(2\gamma)<\tau < L/\gamma$. In all cases, $\tau$ depends on the size, which is compatible with the fact that in the thermodynamic limit we expect an algebraic decay.

We verified these predictions with exact numerical simulations of the full master equation using the stochastic quantum trajectories approach~\cite{Daley_2014} (for $L>4$); we have used the python-based QuTiP package~\cite{Qutip01,Qutip02} that allowed us to push our analysis up to $L=10$ sites with high statistics ($N_{\rm traj}\geq10^3$, $N_{\rm traj}$ being the number of trajectories); whenever not specified, we are using open boundary conditions.
The results of our numerical simulations 
(always with open boundary conditions) 
are shown in Fig.~\ref{Fig:n:BI:LScaling}, where we consider four values of $\hbar\gamma/J$, ranging from $10^{-2}$ to $10$.
Note that here and in the subsequent figures we omitted the error bars, since the statistical errors associated with the averaging over the trajectories remains negligible on the scales of the various plots, up to densities $n \lesssim 10^{-2}$.
For $\hbar\gamma /J\leq 1$ we observe a universal behaviour even at small sizes. 
Data are affected by important finite-size effects and indeed no collapse of curves at different $L$ has been observed (see Appendix~\ref{App:Additional:BI}). 
The comparison with the prediction in~\eqref{Eq:n:BI:TL}, which is plotted as a thin blue dotted line, is satisfactory only at short times and it improves for increasing lattice size. For $L=10$ there is a quantitative agreement until the density decreases to $n\sim 0.4$. 
For longer times, we observe the appearance of an exponential decay, as predicted in the previous paragraph.

We use our numerical simulations to link $\Var N_t$ and $\langle \hat \Pi \rangle_t$ to $N(t)$; motivated by the data shown in Appendix~\ref{App:Additional:BI} we propose:
\begin{subequations}
  \begin{eqnarray}
    \label{Eq:BI:Var:Pi}
    \langle \hat \Pi \rangle_t &\simeq& \frac{N(t)}{L}+\frac{L-2}{4} \frac{N(t)^2}{L^2}, \\
    &&\cr
    \Var N_t &\simeq& 2 \Big[ N(t) - 2 \langle \hat \Pi \rangle_t \Big]  \cr
    &&\cr
    &\simeq& \left(1 - \frac{2}{L} \right) \left( 2N(t) - \frac{ N(t)^2}{L} \right).
  \end{eqnarray}
\end{subequations}
On the other hand, $\langle\hat \Sigma_{\frac \pi 2}\rangle_t$ and $\langle \hat T_u \rangle_t$ are negligible at all times.
We substitute these expressions into Eq.~\eqref{Eq:Dynamical} and obtain the following solution:
\begin{equation}
  n(t)=  \frac{4\left(1-\frac{2}{L}\right)}{L \left(\exp \left[ 
      \left(1- \frac{2}{L} \right) \frac{2\gamma t}{L}
      \right]-1\right)+2-\frac 4L}
  \label{Eq:BI:n:Theory}
\end{equation}
The above equation is the theoretical prediction plotted in Fig.~\ref{Fig:n:BI:LScaling} as a thick dashed blue line, which provides a satisfactory description of our numerics.

\begin{figure}[t]
  \includegraphics[width=0.65\columnwidth]{Test_DE}
  \caption{Time derivative of the population $\dot N(t)$ for $L=8$ and $\hbar \gamma/J = 10^{-1}$. Black solid line: calculation of $\dot N(t)$ using the r.h.s of Eq.~\eqref{Eq:Dynamical} by running a numerical simulation with $N_{\rm traj} = 2000$ quantum trajectories that computes explicitly all the necessary quantities. 
    Red thin line: numerical derivative of the $N(t)$ computed with $N_{\rm traj} = 2000$ quantum trajectories and using the Euler method.}
  \label{Fig:BI:2}
\end{figure}

Since this latter result has been obtained using the numerical data, we perform a direct investigation of whether Eq.~\eqref{Eq:Dynamical} is a good tool for describing the population dynamics.
We compute the numerical derivative of the data displayed in Fig.~\ref{Fig:n:BI:LScaling} for $L=8$ and $\hbar \gamma/J = 0.1$ using Euler's method and we compare the obtained curve with the prediction given by the r.h.s.~of Eq.~\eqref{Eq:Dynamical}, by running a numerical simulation that computes explicitly all the necessary quantities. 
The comparison is shown in Fig.~\ref{Fig:BI:2} and the agreement is excellent, showing that Eq.~\eqref{Eq:Dynamical} models the system even at small sizes. This confirms the general validity of the approximations employed to derive Eq.~\eqref{Eq:Dynamical}.

\subsection{Periodic boundary conditions vs. open boundary conditions}

\begin{figure}[t]
\includegraphics[width=\columnwidth]{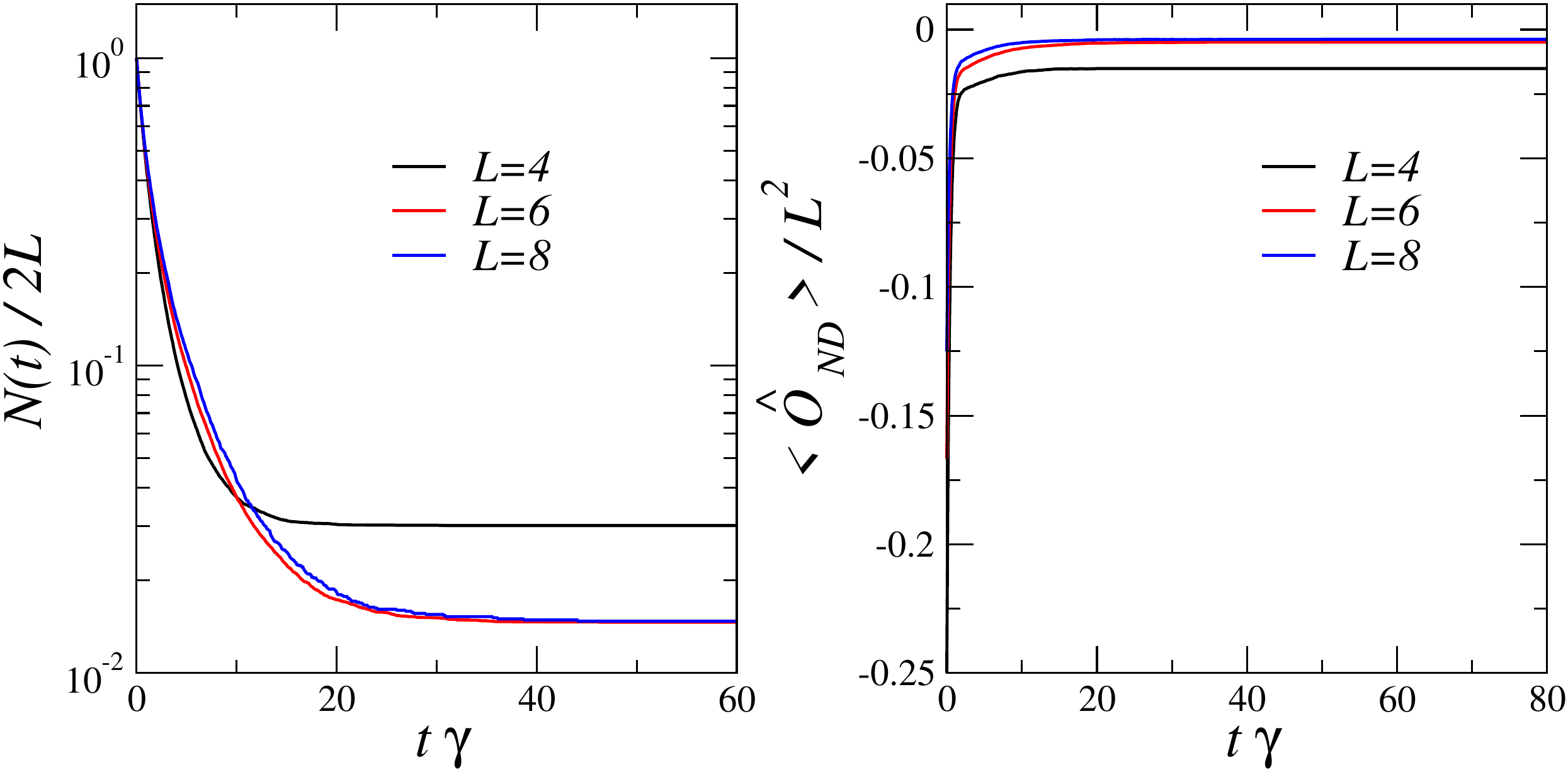}

\vspace{0.3cm}

\includegraphics[width=0.85\columnwidth]{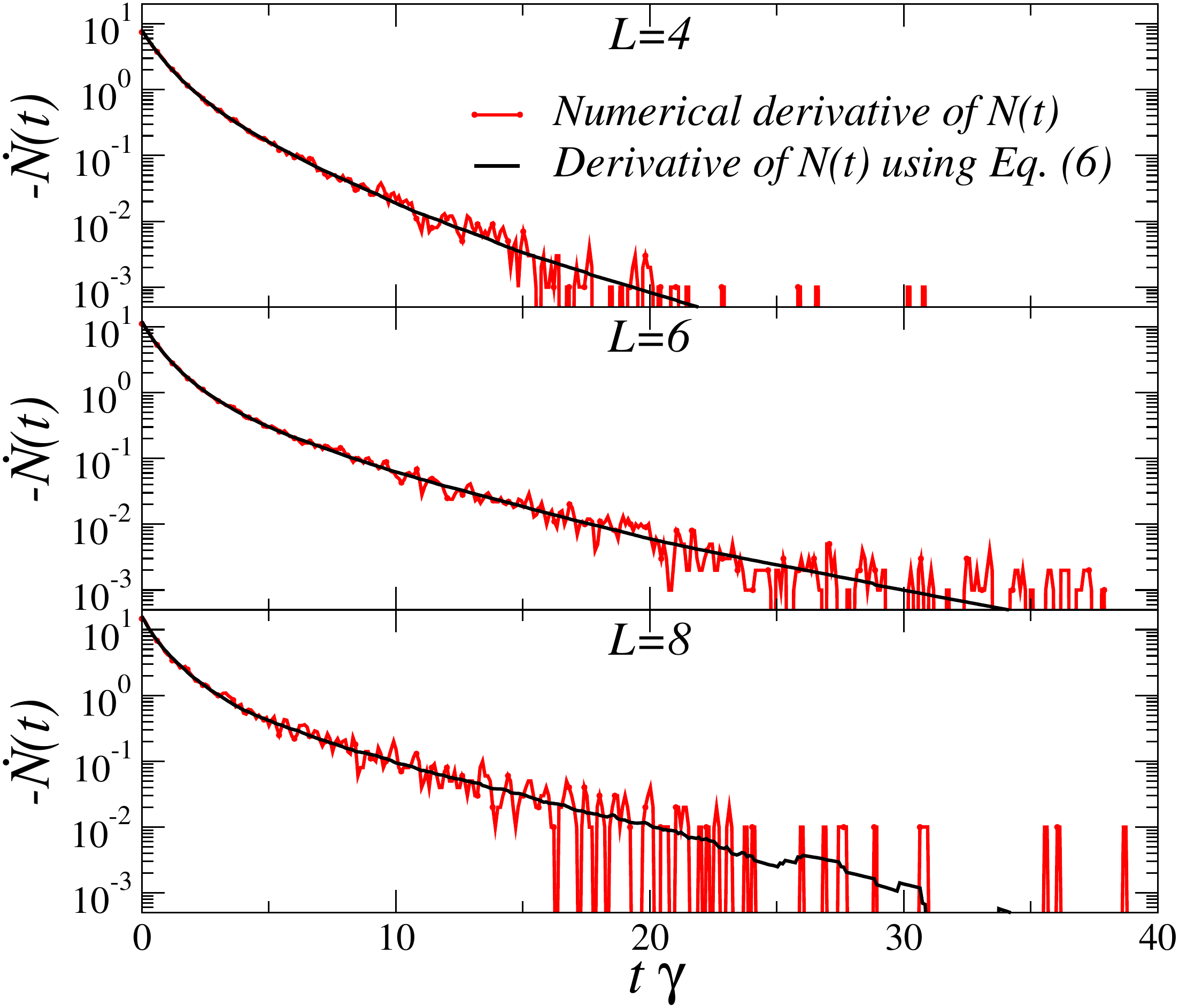}

\caption{Lossy dynamics from an initial band insulator with periodic boundary conditions using quantum trajectories ($N_{\rm traj} = 10^4$ for $L=4,6$ and $N_{\rm traj} = 10^3$ for $L=8$) and $\hbar \gamma / J = 0.1$. Top left: dynamics of the population $N(t)$. Top right: at finite size, the stationary state is not Dicke, as quantified by the operator $\hat O_{\rm ND}$. Bottom panels: time-derivative of the population computed using the r.h.s.~of Eq.~\eqref{Eq:Dynamical} and with the numerical derivative of $N(t)$.}
\label{Fig:BI:pbc}
\end{figure}

The numerical analysis of the dissipative dynamics of an initial band insulator
shows also that for periodic boundary conditions the stationary value of the population is not zero, see Fig.~\ref{Fig:BI:pbc}, top left panel.
This result is in contrast with the expectation that Dicke states are the unique stationary states: since $ \langle \hat S^2 \rangle_0 = 0$, if this were the case, the stationary population should be zero, no matter the boundary conditions.
This problem was also mentioned in Ref.~\cite{FossFeig_2012}.

By looking at Eq.~\eqref{Eq:Dynamical}, we observe that a stationary state that is not a Dicke state is characterised by a non-zero expectation value of the operator $\hat O_{\rm ND}=-\hat \Pi+\hat \Sigma_{\frac \pi2} + \hat T_u $; roughly speaking, $\langle \hat O_{\rm ND}\rangle_{\rm ss}$ measures the \textit{non-Dickeness} of a stationary state.
Whereas our numerics for open boundary conditions shows that $\langle \hat O_{\rm ND}\rangle_t \sim \langle \hat \Pi \rangle_t$ and that $\langle \hat \Pi \rangle_{\rm ss} \sim0$, this is not true for periodic boundary conditions.
In Fig.~\ref{Fig:BI:pbc}, top right panel, we show that in the latter case the stationary value of $\langle \hat O_{\rm ND} \rangle_{\rm ss}$ is different from zero at finite size. This explains why the stationary state is not empty: it is not a Dicke state.

Yet, if we consider $\langle \hat O_{\rm ND} \rangle_{\rm ss} / L^2$ in order to discuss the properties of the thermodynamic limit, such value should tend to zero for $L \to \infty$. We have verified numerically this scaling for $\langle \hat \Sigma_{\frac{\pi}{2}} \rangle_{\rm ss}$ and $\langle \hat T_u \rangle_{\rm ss}$ (not shown). Remarkably, for some values of $L$ the value of $\langle \hat \Pi \rangle_{\rm ss}$ is zero, and for other ones it is not. This absence of a smooth dependence on $L$ prevents us from seeing a clear tendency towards zero for $L \to \infty$ in the plots of $N(t \to \infty) / L$ and $\langle \hat O_{\rm ND} \rangle_{\rm ss} / L^2$ reported in Fig.~\ref{Fig:BI:pbc}. Nonetheless, by mathematical arguments we know that $\langle \hat \Pi \rangle_{\rm ss} / L^2 \to 0$ for $L \to \infty$ because $\hat \Pi$ is the sum of $L$ non-negative and bounded operators. This is sufficient to let us conclude that in the thermodynamic limit $\langle \hat O_{\rm ND} \rangle_{\rm ss} \to 0$ and that the stationary state is Dicke and empty.

We conclude this subsection with two final messages. First, Eq.~\eqref{Eq:Dynamical:TL} retains its validity even with periodic boundary conditions, when numerical simulations show finite-size effects that qualitatively deviate from the assumption that Dicke states are the only stationary states.
Second, even in these situations, at finite size, Eq.~\eqref{Eq:Dynamical} can be employed to describe the dynamics, and to characterise deviations from stationary Dicke states. In order to support the latter statement, in Fig.~\ref{Fig:BI:pbc}, bottom panel, we compare the time derivative of the population computed from the numerical value of $N(t)$ and from Eq.~\eqref{Eq:Dynamical}, displaying an excellent agreement.

\section{Dynamics from a Mott insulator}
\label{Sec:Mott}

\subsection{Thermodynamic limit}

We now consider an initial state with one particle per site, that is, a Mott insulator in the atomic limit.

 Because of the spin, the manifold of such states spans a subspace of dimension $2^L$.
Conservation of spin during the dynamics takes here a non-trivial form because the spin of the gas can range from $0$ to $L/2$. 
We can easily discuss the dynamics in the thermodynamic limit: the asymptotic number of particles is $n_{\infty} = \frac{2}{\hbar} s_0$ and the dynamics is given by:
\begin{equation}
  n(t) =  n_{\infty} \tanh \left[ \frac {n_{\infty} \gamma } 2 t + \text{arctanh} \left( \frac 1{n_{\infty}} \right)\right].
  \label{Eq:MOTT:TL}
\end{equation}
This result displays in a clear way the interplay between spin-conservation and dissipative dynamics. Not only the stationary properties of the gas are determined by the initial spin of the gas: the dynamics too is determined by it, since stationary properties are approached with a typical decay time 
\begin{equation}
  \tau = \frac{2}{\gamma n_{\infty}} = \frac{\hbar}{\gamma s_0}
  \label{Eq:tau:ninf:spin}
\end{equation} 
that depends on spin, and that is shorter for larger spin values.

The link between the stationary number of particles and the typical decay time has already been highlighted in Ref.~\cite{Sponselee_2018}, although the authors do not mention its connection with spin.

\subsection{Finite-size effects}

If we consider a Mott insulator with $N=L$ particles initialised in an eigenstate of $\hat S^2$ with quantum number $S$, the asymptotic number of particles can be exactly characterised for any size assuming that the final state is a Dicke state. In this case we expect that the stationary state has a well defined number of particles given by the relation $N_{\infty}=2S$. 

When the initial Mott insulator is not an eigenstate of $\hat S^2$, we expect the final state to be a linear superposition of Dicke states with different number of particles ($\Var N > 0$) and spin. Assuming stationarity, $\dot N=0$, and recalling that for a linear superposition of Dicke states $\langle \hat \Pi \rangle = 0$, we obtain:
\begin{equation}
  N_{\infty} \leq \sqrt{\frac{4 \langle \hat S^2\rangle}{\hbar^2}+1}-1.
  \label{Eq:UpperBound}
\end{equation}
It is simple to verify that the bound is saturated when the initial state is an eigenstate of $\hat S^2$.

\subsection{Typical decay time and spin}

\begin{figure}[t]
  \includegraphics[width=\columnwidth]{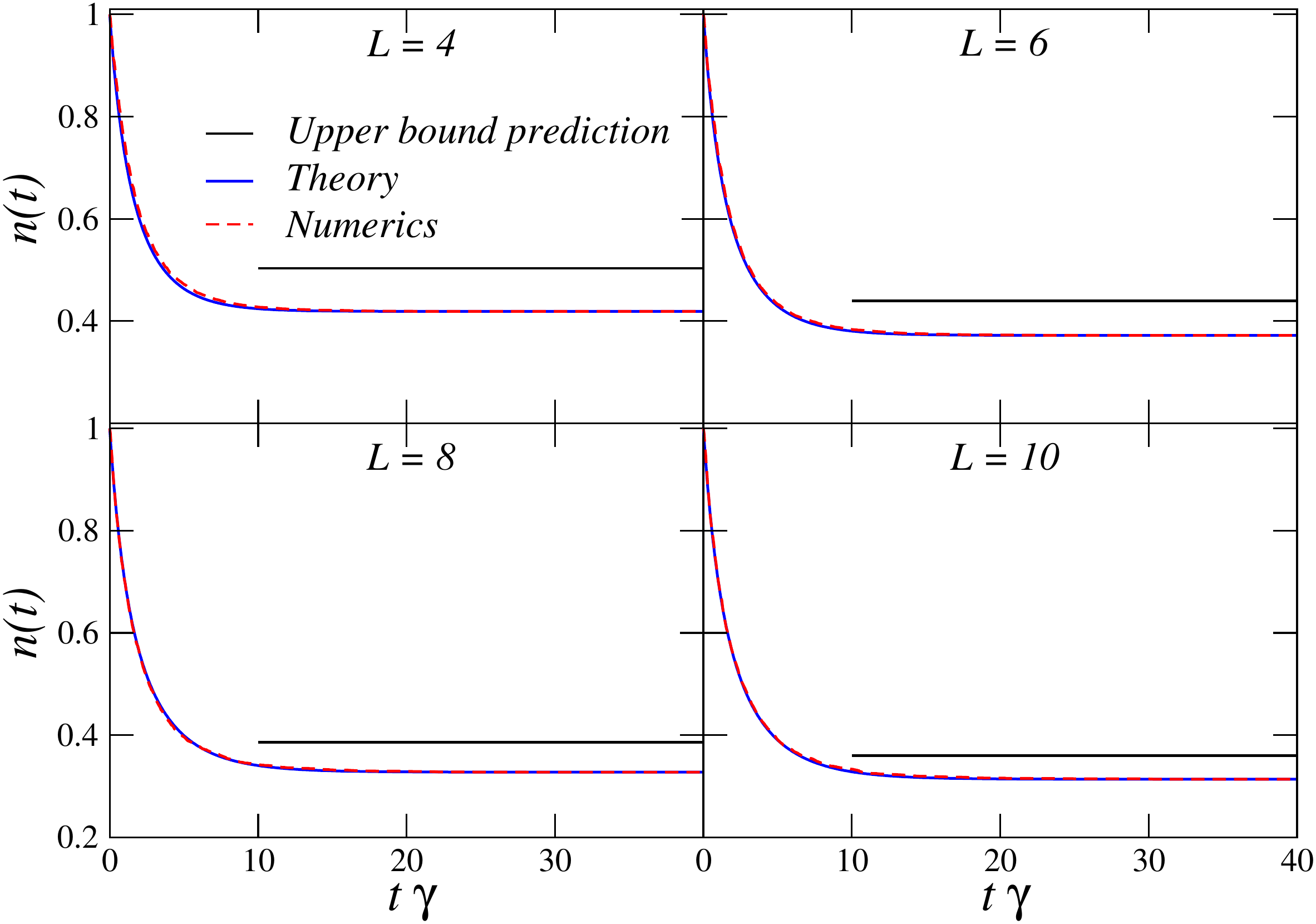}
  \caption{Dissipative dynamics of a N\'eel state for $L=4$, $6$, $8$ and $10$ for $\hbar \gamma/J = 0.1$. The time-dependent population is plotted as a dashed red line. The upper bound to the total population given in Eq.~\eqref{Eq:UpperBound} is in black. The blue curve is a fit to the dynamics using Eq.~\eqref{Eq:MOTT:TL} and taking $n_{\infty}$ as the only fitting parameter.}
  \label{Fig:Neel}
\end{figure}

In order to test these predictions we performed several numerical simulations of the full master equation for an initial Mott insulator. We consider as initial state a N\'eel state with alternating spin up and down: $\ket{\uparrow \; \downarrow \; \uparrow\; \downarrow \ldots}$. Results for the decaying populations are shown in Fig.~\ref{Fig:Neel}.
We observe that, since the initial state is not an eigenstate of $\hat S^2$, the bound of Eq.~\eqref{Eq:UpperBound} is satisfied but not saturated.
It is shown that in this situation Eq.~\eqref{Eq:MOTT:TL} provides an excellent description of the dynamics taking only $n_{\infty}$ as fit parameter. We have verified that the fitted value of $n_{\infty}$ does not satisfy the relation with the spin, i.e.~$n_{\infty} \ne \frac{2 \sqrt{\langle \hat S^2 \rangle}}{\hbar L} $.

\begin{figure}
  \includegraphics[width=\columnwidth]{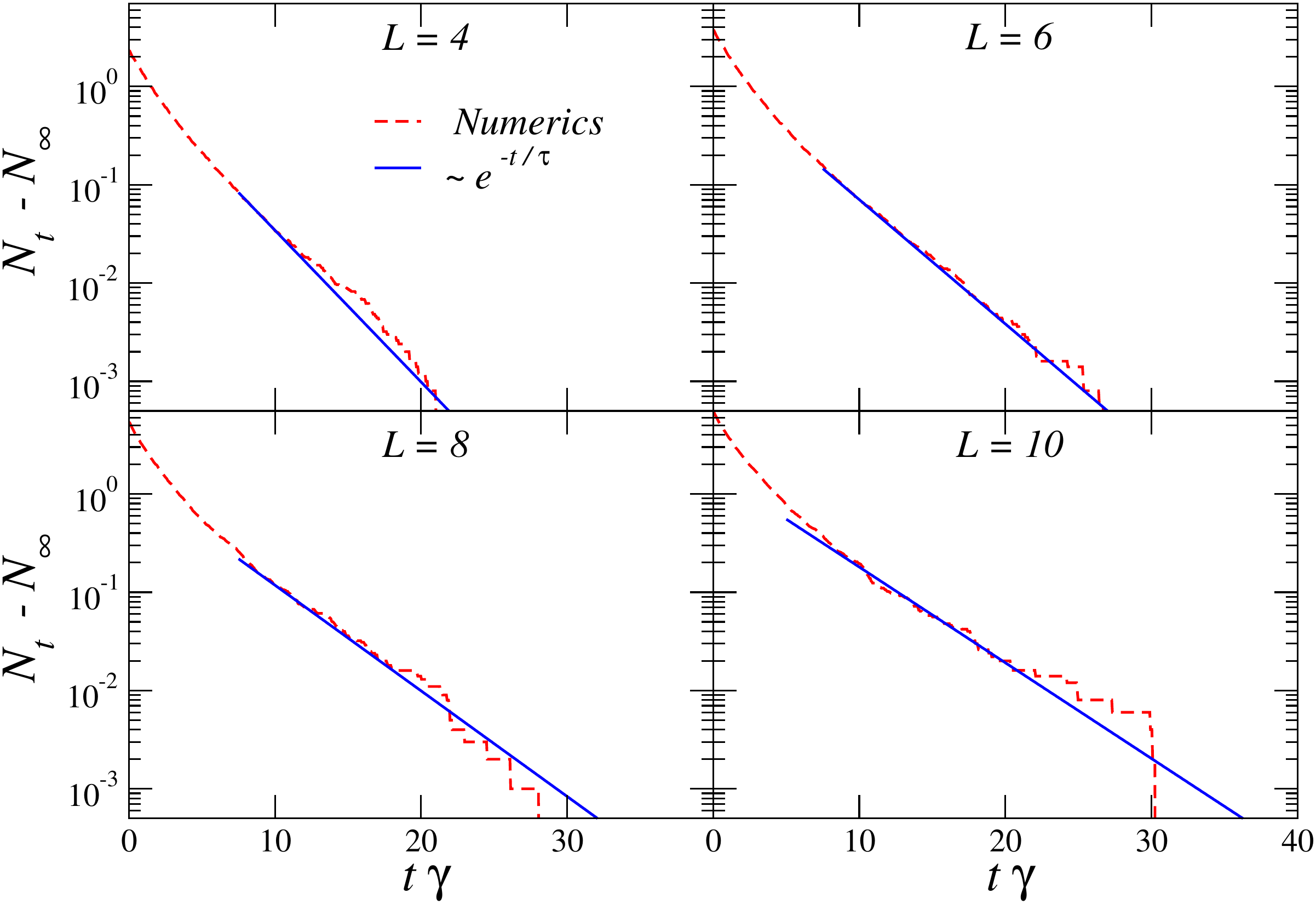}
  \caption{Dissipative dynamics of a N\'eel state for $L=4$, $6$, $8$ and $10$  for $\hbar \gamma/J = 0.1$ and long-time decay of $N(t)$ to the stationary value $N_{\infty}$. In all cases we observe an exponential decay. The blue curve is not a fit, but the decay obtained from the theoretical prediction $N_{t}-N_{\infty}=e^{-t/\tau}$ with $\tau$ given by Eq.~\eqref{tauestim}, which provides an excellent description of the decay time at all lattice lengths.}
  \label{Fig:Tau:Spin}
\end{figure}

We further investigate the typical time-scale with which the asymptotic number of particles is approached. 
The results in Fig.~\ref{Fig:Tau:Spin} show a clear exponential approach to the stationary value. We have investigated whether the analytical formulas given in Eq.~\eqref{Eq:tau:ninf:spin} can give a quantitative prediction to the typical decay time $\tau$. As shown in the plot, the formula 
\begin{equation}
  \label{tauestim}
  \tau = \frac{\hbar L}{\gamma \sqrt{\langle \hat S^2 \rangle}}
\end{equation}
gives a remarkably good description of the numerical data. Less accurate results are instead obtained with the formula $\tau = 2/(\gamma n_{\infty})$, where $n_{\infty}$ has been taken from the previous fit; the discrepancy is solely ascribed to finite-size effects and is expected to disappear in the thermodynamic limit. 

We finally perform a set of numerical simulations to test whether this latter relation between the decay time and the spin is true in general. We consider as initial state an uncorrelated Mott insulator where, on each lattice site, the spin $\vec S_i$ is randomly oriented. 
In order to construct such a state, we randomly draw two angles, $\theta_i$ and $\phi_i$, which identify a generic direction on the Bloch sphere, for every lattice site $i$. The goal of this procedure is not to sample in a uniform way the set of uncorrelated Mott insulators, but to generate states with widely variable values of $\langle \hat S^2 \rangle$.
We have evolved in time 14 Mott insulators with uncorrelated and random site-dependent spin alignment; in all cases we observe an exponential approach to the stationary value of the number of particles.

\begin{figure}[t]
  \includegraphics[width=0.67\columnwidth]{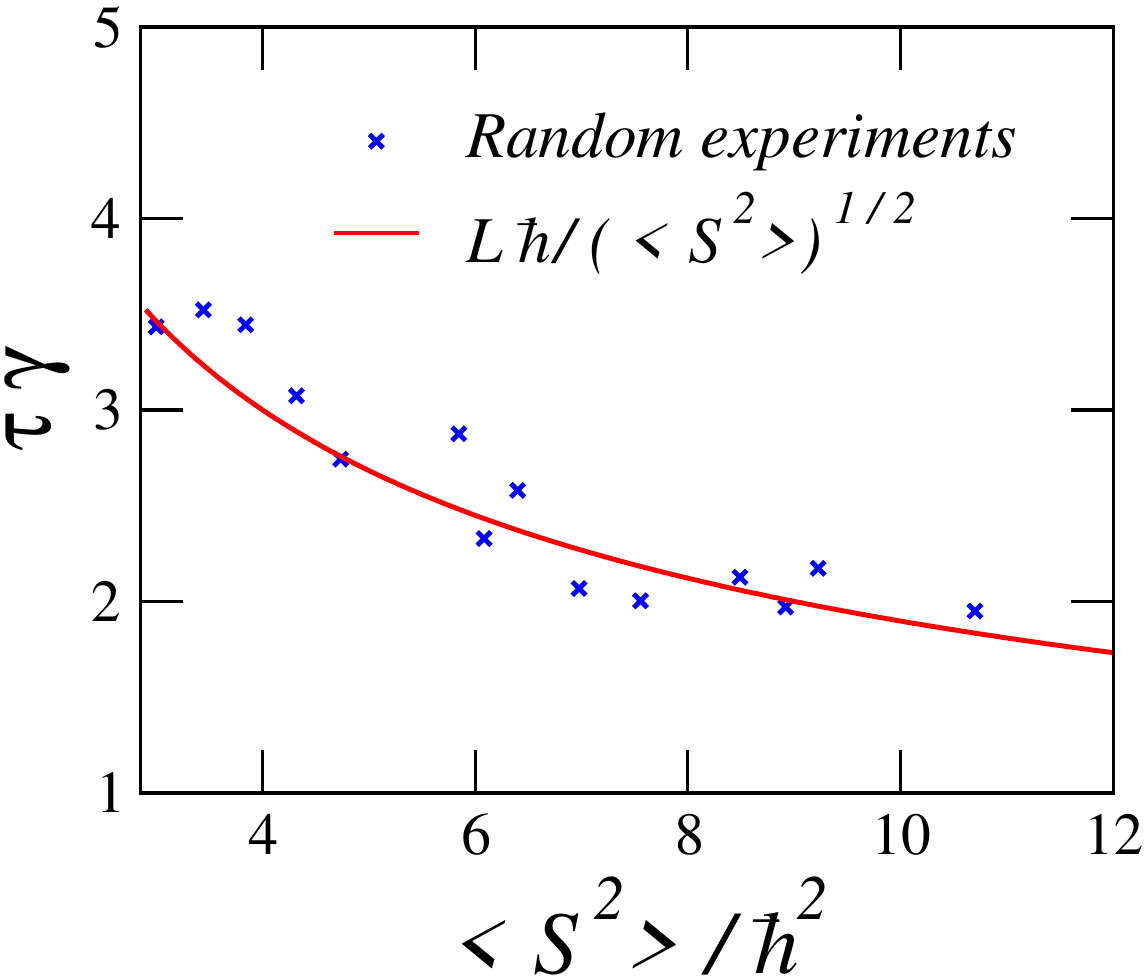}
  \caption{Decay time of the population $N(t)$ approaching the asymptotic number of particles $N_{\infty}$ as a function of $\langle S^2 \rangle/\hbar^2$, for $L=6$ and $\hbar \gamma /J= 0.1$. Red solid line: theoretical curve for $\tau$ predicted in the thermodynamic limit \eqref{tauestim}. Blue crosses: numerical fits of the decay time $\tau$ performed for 14 ``random" Mott insulators with $10^4$ quantum trajectories.}
  \label{Fig:MI:Rand}
\end{figure}

We fit the typical time-scale with which the asymptotic value is approached and compare it to the theoretical prediction given by Eq.~\eqref{tauestim}. 
We first take as $N_{\infty}$ the numerical value of the population at the longest computed time; we then fit $N(t) - N_{\infty}$ at intermediate times, because at long times its value is comparable with the statistical error bars due to our stochastic sampling with quantum trajectories. The results shown in Fig.~\ref{Fig:MI:Rand} display a clear correlation between the typical decay time predicted by the theory and the fitted one. Notice that the theoretical prediction has been derived in the thermodynamic limit, whereas here we consider numerical simulations for $L=6$.

\subsection{Symmetry-resolved purity}
\label{Sec:Symm:Res:Pur}

When starting from a Mott insulator, the system features a non-trivial dynamics also in terms of the purity of the total density matrix: 
\begin{equation}
  \label{purdef}
  \mathcal{P}_{\rm tot} = \textrm{tr}\left[{\rho}(t)^2\right].
\end{equation}
Since the density matrix is reconstructed via the independent dynamics of $N_{\rm traj}$ quantum trajectories~\cite{Daley_2014}, we have that 
\begin{equation}
  \label{rhotraj}
  {\rho}(t) = \frac{1}{N_{\rm traj}}  \sum_{i=1}^{N_{\rm traj}}  \ket{\psi_i(t)} \hspace{-0.06cm} \bra{\psi_i(t)},
\end{equation}
where $\ket{\psi_i(t)}$ is the $i$-th trajectory at time $t$. 
Thus, combining Eq.~\eqref{rhotraj} with Eq.~\eqref{purdef} we get (from now on we will omit the time dependence)
\begin{align}
  \label{tr}
  \mathcal{P}_{\rm tot} =& \frac{1}{N_{\rm traj}^2} \sum_{s} \sum_{i,j=1}^{N_{\rm traj}} \langle s | \psi_i\rangle  \langle\psi_i | \psi_j\rangle  \langle\psi_j | s\rangle \cr
  =& \frac{1}{N_{\rm traj}^2} \sum_{i,j=1}^{N_{\rm traj}} \left|\langle\psi_i | \psi_j\rangle\right|^2
\end{align}
where $\left\{\ket{s}\right\}$ is an orthonormal basis of the Hilbert space and in the second line we used that $\sum_s\ket{s} \hspace{-0.06cm} \bra{s}= \mathbb{I}$.

\begin{figure}[!t]
  \includegraphics[width=0.7\columnwidth]{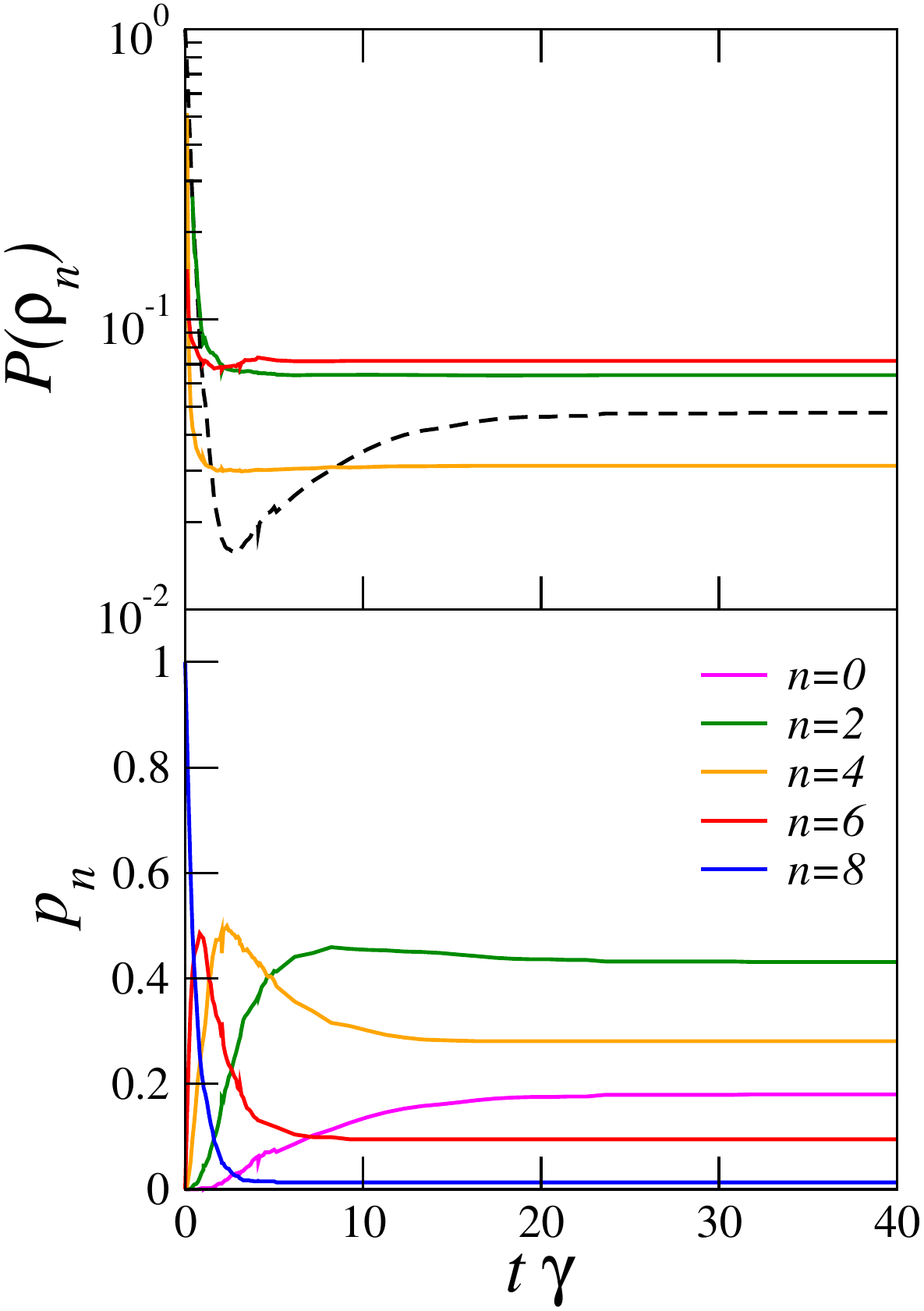} 
  \caption{Symmetry-resolved purity for the particle sectors $n=2,4,6$ (solid lines) and purity of the full density matrix (dashed line). Data are obtained for a typical set of parameters $J/\gamma=10\hbar$, $L=8$, and $N_{\rm traj}=10^3$.}
  \label{Fig:purity_mott_symres}
\end{figure}

The dynamics of a single quantum trajectory
is governed by an effective non-Hermitian Hamiltonian $\hat{H}_{\rm eff} = \hat{H}-i(\gamma/2)\sum_{i=1}^L\hat{L}^\dagger_i\hat{L}_i$ and by the stochastic  quantum jumps determined by the jump operators $\hat{L}_i$~\cite{Daley_2014}.
While the evolution induced by $\hat{H}_{\rm eff}$ conserves the number of particles, the quantum jumps do not: they couple the $n$- with the $(n-2)$-particle sector of the Hilbert space.
For this reason, if the initial state of is an eigenstate of $\hat{N}$, each quantum trajectory $\ket{\psi_i(t)}$ will have at any time a well defined (although time dependent) number of particles.  
We can thus label the trajectories with a double index, $\ket{\psi_{n,\alpha}}$, where $n$ is the particle sector and $\alpha$ labels the trajectories belonging to the $n$-th subspace. Note that $n$ depends on time.

Using the fact that $\langle\psi_{n,\alpha} | \psi_{m,\beta}\rangle = 0$ for $n\neq m$ we write: 
\begin{equation}
  {\rho}= \bigoplus_n p_n \rho_n \,,
  \qquad p_n=\frac{N_n}{N_{\rm traj}} \,,
\end{equation}
where $N_n$ is the number of trajectories belonging to the $n$-particle sector and
\begin{equation}
  \rho_n = \frac{1}{N_n} \sum_{\alpha=1}^{N_n} \ket{\psi_{n,\alpha}} \hspace{-0.06cm} \bra{\psi_{n,\alpha}}, \quad \text{ with } \; {\rm tr}\left[\rho_n\right]=1.
\end{equation}
We can thus link the total purity $\mathcal P_{\rm tot}$ to the symmetry-resolved purities $\mathcal P(\rho_n)$, i.e.~the purities of the symmetry-resolved density matrices:
\begin{equation}
  \mathcal{P}_{\rm tot} =\sum_{n} p_n^2  \ \mathcal{P}\left(\rho_n\right).
\end{equation}

We now study the time-evolution of the total purity and of the symmetry-resolved purities for an initial Mott insulator with N\'eel order; we perform numerical simulations for $L=8$ and $\hbar \gamma/J = 0.1$.
In Fig.~\ref{Fig:purity_mott_symres} we show the dynamics of the symmetry-resolved purities for the  sectors $n=2$, $4$ and $6$ (the purities for $n=8$ and $n=0$ are trivial and equal $1$) and the behaviour of the occupation probabilities $p_n$.
The plot of $\mathcal{P}_{\rm tot}$ (dashed line in the top panel) shows
that for $n=2$ and $6$ the symmetry-resolved purity is larger than the total one at long times, i.e.~$\mathcal{P}\left(\rho_n\right)>\mathcal{P}_{\rm tot}$. 

\begin{figure}[!t]
  \includegraphics[width=0.7\columnwidth]{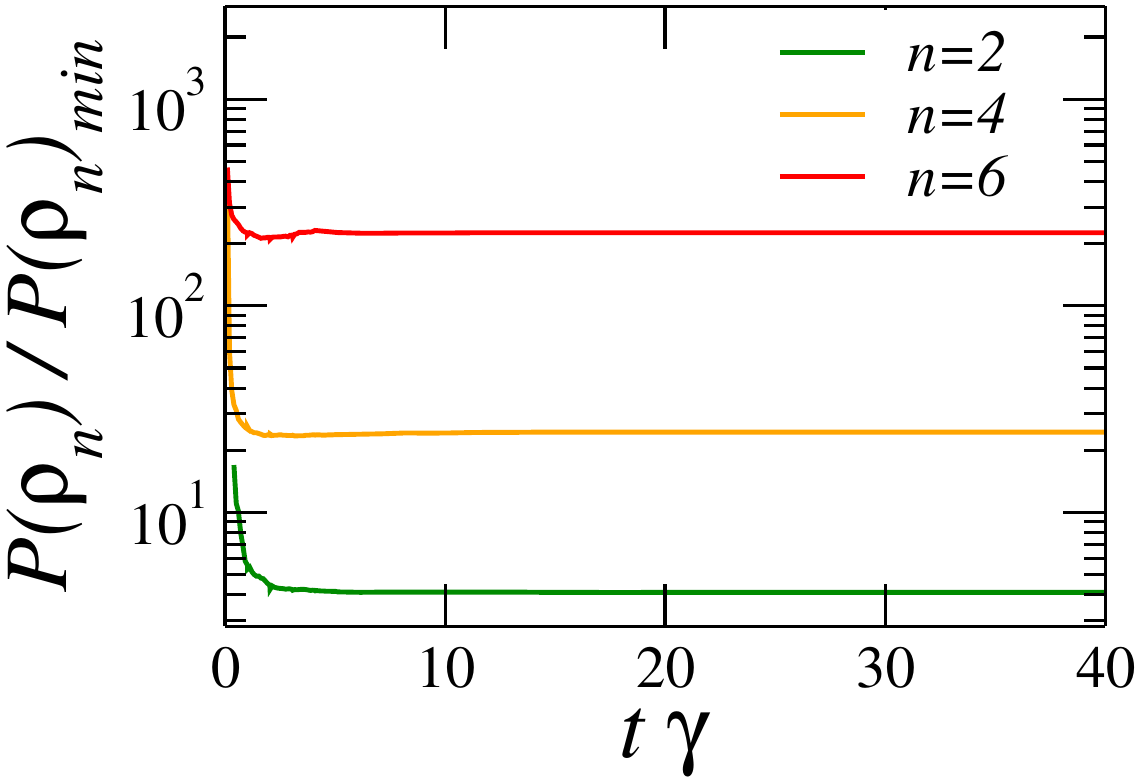} 
  \caption{Purity ratio $\mathcal{P}\left(\rho_n\right)/\mathcal{P}\left(\rho_n\right)_{\rm min}$ for $n=2,4,6$. Parameters are set as in Fig.~\ref{Fig:purity_mott_symres}.}
  \label{Fig:purity_mott_ratio}
\end{figure}

In Fig.~\ref{Fig:purity_mott_ratio} we show the symmetry-resolved purity normalised by its minimum possible value for a given $n-$particle subspace, i.e. the purity of a fully mixed state $\mathcal{P}\left(\rho_n\right)_{\rm min}=1/{\rm dim}(\mathcal{H}_n)$, where ${\rm dim}(\mathcal{H}_n)=\binom{L}{n/2}^2$ is the dimension of the $n$-particle subspace of the Hilbert space (with $S_z=0$).
Remarkably, the asymptotic dynamics features purities which are larger by orders of magnitude with respect to  $\mathcal{P}\left(\rho_n\right)_{\rm min}$. 
This is related to the fact that, despite our system is subject to particle losses, the non-trivial interplay between spin conservation and dissipation leads to the creation of a non-trivial dark subspace for all possible number of particles $n$.
It is interesting to observe that differently from what reported in Ref.~\cite{vitale2021} the purification process here is not transient, but takes place in the long-time limit.

\section{The effect of weak interactions}
\label{Sec:Interactions}

\begin{figure}[!t]
  \includegraphics[width=0.67\columnwidth]{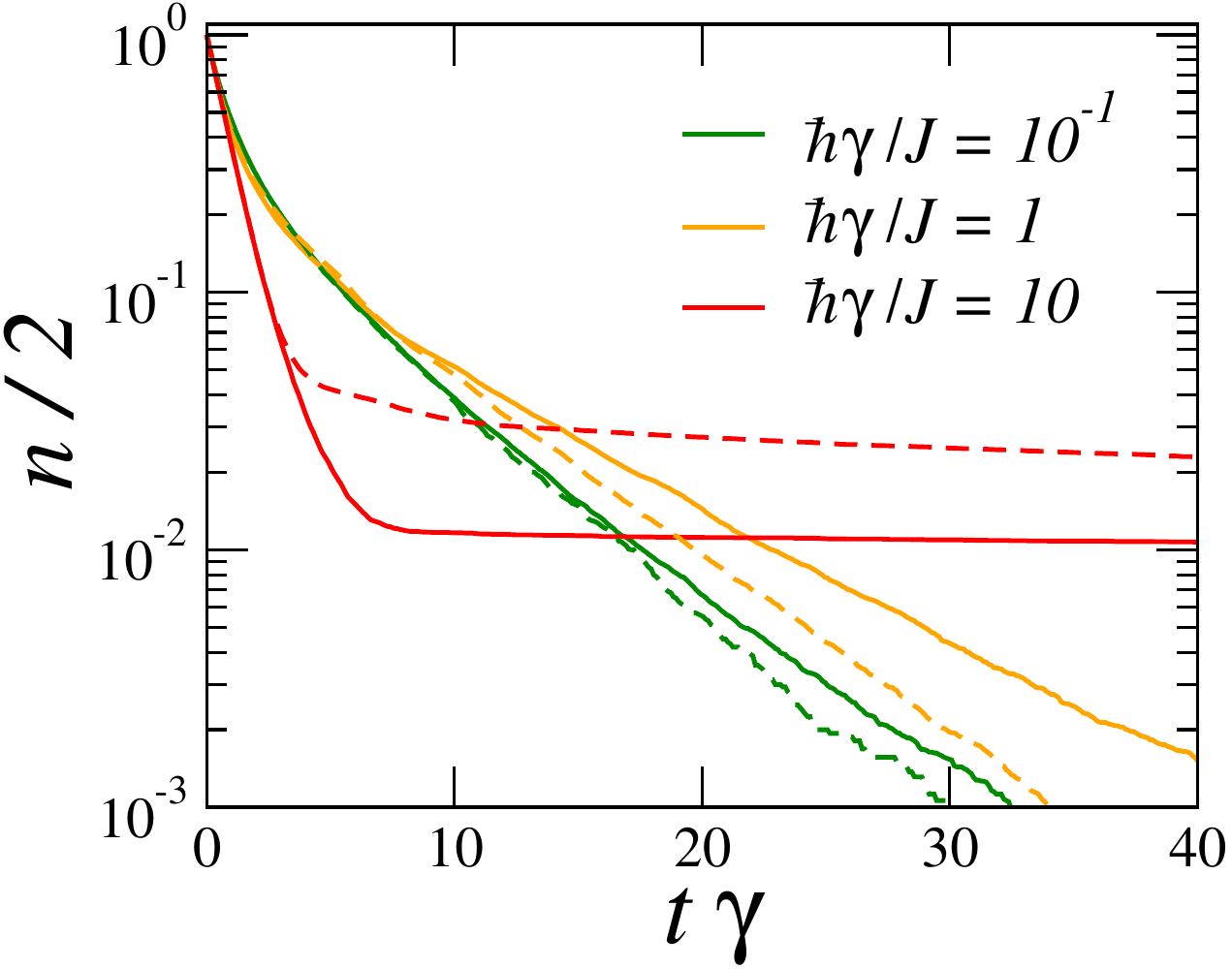}
  \caption{Dissipative dynamics of the normalised density of a band insulator for $L=8$ and $10^4$ quantum trajectories. Different colors refer to different dissipation strengths, from $\hbar\gamma/J=0.1$ to $\hbar\gamma/J=10$. Solid lines: simulations with $U= \hbar \gamma$. Dashed lines: corresponding dynamics for $U=0$. The plot highlights the collapse of the curves for $U \ne 0$ and $U = 0$ in the weakly dissipative limit.}
  \label{Fig:BI:Weak}
\end{figure}

We now discuss the validity of the approximation introduced in Sec.~\ref{Sec:Pop_dynamics}, concerning the complete neglection of the interaction term in the Hubbard Hamiltonian. The discussion presented in this article focuses on the limit of weak dissipation $\hbar \gamma \ll J$, that in most experimental situations coincides with the limit of weak interactions $U \ll J$.
In Fig.~\ref{Fig:BI:Weak} we show numerical simulations performed with a finite value of the interaction constant, $U=\hbar \gamma $, starting from a band insulator. The numerics clearly shows that the presence of interactions does not affect significantly the dynamics, when they are weak. The curious thing is that a weak dissipation, instead, can significantly affect the dynamics.

The reason why we can safely neglect interactions but not dissipation lies in the separation of time scales between unitary hopping dynamics (very fast) and dissipative/interaction dynamics (slow). Our theoretical analysis is a perturbative treatment of dissipation, and the population equation~\eqref{Eq:Dynamical} is linear in $\gamma$; it can be regarded as a first-order expansion of the correct equation. 
It is well known from the standard time-dependent perturbation theory and Fermi golden rule that a perturbative treatment of the unitary evolution due to interactions gives transition rates that are of order $U^2$, and thus negligible with respect to the dissipative dynamics.
Since $\hbar \gamma$ and $U$ are of the same order of magnitude,
we obtain that interactions give a second-order correction in the weakly dissipative limit we deal with. 
For this reason, the whole discussion, performed in the $U=0$ limit, is expected to provide accurate results for the weakly dissipative/interacting limit.

\section{Conclusions}
\label{Sec:Concl}

We have presented a theoretical study of the dissipative dynamics of a one-dimensional fermionic gas subject to two-body losses. Our study has focused on the non-trivial interplay between dissipation and spin conservation, as highlighted by the differential equation obeyed by the gas population, see Eq.~\eqref{Eq:Dynamical}.
Not only we have shown that the stationary population is due to the initial spin of the gas, but also that the dynamics and its typical asymptotic decay time depends on the spin. Our analysis has focused on three kinds of initial states (Dicke states, band insulator and several kinds of Mott insulators), characterised by different dynamical properties.

The simplicity of the proposed equations, especially when considered in the thermodynamic limit, makes us think that they could have application in the modelisation of experimental studies~\cite{Yan_2013, Zhu_2014, Sponselee_2018}. Although the experiments have so far mainly focused on the strongly-dissipative Zeno limit, the fact that similar equations have been phenomenologically used for fitting experimental data~\cite{Sponselee_2018} hints at a possible use also in this regime (after replacing $\gamma$ with the Zeno decay rate).
The generalisation of our results to the strongly-dissipative case~\cite{Rossini_2020} remains the most important perspective of this study.
Furthermore, the possibility to stabilize Dicke states with a non-trivial entanglement content opens the exciting possibility to characterize the entanglement properties of the system, analogously to what has been done in Ref.~\cite{Goto_2020} for bosonic particles, with also possible applications to quantum metrology~\cite{Apellaniz2015}.

\acknowledgements

We thank the Hamburg team working on Yb gases (B.~Abeln, C.~Becker, M.~Diem,  T.~Petersen and K.~Sponselee) for stimulating discussions. We also acknowledge enlightening discussions with O.~Giraud, F.~Minganti and B.~Vermersch.
This work has been partially funded byLabEx PALM (ANR-10-LABX-0039-PALM).

\appendix

\section{Derivation of Eq.~\eqref{Eq:Dynamical}}
\label{App:Derivation:Dyn}

We start from Eq.~\eqref{Eq:Expectation:NN} and focus on $\langle \hat c^\dagger_{k, \uparrow} \, \hat c_{w, \uparrow} \, \hat c^\dagger_{q, \downarrow} \, \hat c_{z, \downarrow} \rangle_t$. We retain only the momentum-conserving ($k+q=w+z+2 \pi n$) and energy-conserving ($\omega_k+\omega_q=\omega_w+\omega_z$, with $\hbar \omega_k = -2 J \cos k$) correlators. 
We have identified five possibilities and the first three read:
\begin{itemize}
  \item $k=q=w=z$. In this case: $$\langle \hat c^\dagger_{k, \uparrow} \, \hat c_{w, \uparrow} \, \hat c^\dagger_{q, \downarrow} \, \hat c_{z, \downarrow} \rangle_t = \langle \hat n_{k, \uparrow} \, \hat n_{k,\downarrow}\rangle_t.$$
  \item $k=w$, $q=z$ and $k \neq q$. In this case: $$\langle \hat c^\dagger_{k, \uparrow} \, \hat c_{w ,\uparrow} \, \hat c^\dagger_{q, \downarrow} \, \hat c_{z, \downarrow} \rangle_t = \langle \hat n_{k \uparrow} \, \hat n_{q,\downarrow}\rangle_t \quad \text{with} \; k \neq q.$$
  \item $k=z$, $q=w$ and $k \neq q$. In this case: $$\langle \hat c^\dagger_{k, \uparrow} \, \hat c_{w, \uparrow} \, \hat c^\dagger_{q, \downarrow} \hat c_{z, \downarrow} \rangle_t = \langle \hat c^\dagger_{k, \uparrow} \, \hat c_{k,\downarrow} \, \hat c^\dagger_{q,\downarrow} \, \hat c_{q,\uparrow}\rangle_t \quad \text{with} \; k \neq q.$$
\end{itemize}

If we only consider these processes, we obtain:
\begin{equation}
  \langle \hat \Pi \rangle_t + \sum_{k \neq q} \langle \hat n_{k, \uparrow} \hat n_{q,\downarrow}\rangle_t + \langle \hat c^\dagger_{k, \uparrow} \hat c_{k,\downarrow} \hat c^\dagger_{q,\downarrow} \hat c_{q,\uparrow}\rangle_t.
\end{equation}
We simplify this expression by introducing the explicit expressions for the the spin operators. From the relations:
\begin{equation}
  \sum_{k} \hat n_{k \uparrow, \downarrow} =  \frac{\hat N}{2} \pm \frac{\hat S_z}{\hbar} \,, \quad
  \sum_k \hat c^\dagger_{k \uparrow} \, \hat c_{k\downarrow}=
  \frac{\hat S_x+ i \hat S_y}{\hbar}\,;
  \label{Eq:Spin:nk}
\end{equation}
one easily obtains that:
\begin{subequations}
  \begin{eqnarray}
    \frac{\hat N^2}{4}-\frac{\hat S_z^2}{\hbar^2} & = & \sum_{k,q}  \hat n_{k, \uparrow} \, \hat n_{q, \downarrow} \,; \\
    \frac{\hat S_x^2 + \hat S_y^2 + \hbar \hat S_z}{\hbar^2} & = & \sum_{kq} \hat c^\dagger_{k, \uparrow} \, \hat c_{k,\downarrow} \, \hat c^\dagger_{q,\downarrow} \, \hat c_{q,\uparrow} \,.
  \end{eqnarray}
\end{subequations}
Having the care of splitting the sums $\sum_{k,q}$ as $\sum_{k=q}+\sum_{k \neq q}$ one obtains the first five terms of Eq.~\eqref{Eq:Dynamical}.

We now consider the latter two terms
\begin{itemize}
 \item $q = \pi-k$, $z = \pi -w$ and $k \neq w$ and $k \neq \pi -w$. In this case:
 $$
 \langle \hat c^\dagger_{k, \uparrow} \, \hat c_{w, \uparrow} \, \hat c^\dagger_{q, \downarrow} \hat c_{z, \downarrow} \rangle_t = \langle \hat c^\dagger_{k, \uparrow} \, \hat c_{w,\uparrow} \, \hat c^\dagger_{\pi-k,\downarrow} \, \hat c_{\pi-w,\downarrow}\rangle_t \quad \text{with} \; k \neq w,\pi-w.
 $$
 It describes processes that are symmetric with respect to lattice momentum $\pi/2$ and all momenta appearing in this expression should be intended $\mod 2 \pi$, so that they can be restricted to the first Brillouin zone $[- \pi, \pi]$. It is responsible for the term $\hat \Sigma_{\frac \pi2}$ in Eq.~\eqref{Eq:Dynamical}.
 \item Finally we have to consider umklapp processes, where momentum is conserved modulus $2 \pi$. Two classes of processes transferring momentum $+2 \pi $ are possible:
 \begin{itemize}
  \item[$\circ$] $w = -k$, $z = -q$ and $k+q = \pi$;
  \item[$\circ$] $w = -q$, $z = -k$ and $k+q = \pi$. 
 \end{itemize}
 Two similar opposite processes are possible that transfer momentum $- 2 \pi$. These processes are responsible for the term $\hat T_{u}$ in Eq.~\eqref{Eq:Dynamical}.
\end{itemize}

\section{Gaussian density matrix: a dynamical equation for the thermodynamic limit}\label{Sec:Gaussian}

We start from Eq.~\eqref{Eq:Expectation:NN} and focus on $\langle \hat c^\dagger_{k, \uparrow} \, \hat c_{w, \uparrow} \, \hat c^\dagger_{q, \downarrow} \, \hat c_{z, \downarrow} \rangle_t$. We now assume that the density matrix is Gaussian and that Wick's theorem applies: 
\begin{align}
 \langle \hat c^\dagger_{k, \uparrow} \, \hat c_{w, \uparrow} \, \hat c^\dagger_{q, \downarrow} \, \hat c_{z, \downarrow} \rangle_t \sim & \: \langle \hat c^\dagger_{k, \uparrow} \, \hat c_{w, \uparrow} \rangle_t \: \langle \hat c^\dagger_{q, \downarrow} \, \hat c_{z, \downarrow} \rangle_t+ \nonumber \\&- \langle \hat c^\dagger_{k, \uparrow} \, \hat c_{z, \downarrow} \rangle_t \: \langle \hat c^\dagger_{q, \downarrow} \, \hat c_{w, \uparrow} \rangle_t .
\end{align}
Of all the correlators which appear here, we only retain those which do not have an explicit time dependence because dissipation is weak and they average to zero between two dissipative events. Thus:
\begin{equation}
 \dot N(t) = -\frac{2 \gamma}{L} \sum_{k,q} \left( \langle \hat n_{k, \uparrow} \rangle_t \:\langle \hat n_{q, \downarrow} \rangle_t - \langle \hat c^\dagger_{k, \uparrow} \hat c_{k, \downarrow}\rangle_t \: \langle \hat c^\dagger_{q, \uparrow} \hat c_{q, \downarrow}\rangle_t \right).
\end{equation}
By using the expressions~\eqref{Eq:Spin:nk}, we finally obtain the dynamical equation in~\eqref{Eq:Dynamical:TL}.

\section{Proof of relations~\eqref{eq:Dicke} for Dicke states}
\label{App:Dicke:Relations}

In this appendix we prove the relations~\eqref{eq:Dicke1} and~\eqref{eq:Dicke2} that characterise Dicke states listed in Sec.~\ref{Sec:Dicke}.

Relation~\eqref{eq:Dicke1} is simple and follows from the definition of Dicke state: $\bra{\Psi_{\rm D}} \hat S^2 \ket{\Psi_{\rm D}} = \hbar^2 \sum_N |c_N|^2 \frac{N}{2} \left(\frac{N}{2}+1 \right) = \frac{\hbar^2}{4}\langle \Psi_{\rm D} |\hat N^2 |\Psi_{\rm D}\rangle + \frac{\hbar^2}{2} \langle \Psi_{\rm D} | \hat N | \Psi_{\rm D} \rangle$.

We begin by considering the the former of the relations in~\eqref{eq:Dicke2}.
We first demonstrate it for a generic Dicke state with well-defined number of particles $N$ and that is also an eigenstate of $\hat S_z$ with eigenvalue $\hbar m$,  $\ket{{\rm D}_{N,m}}$.
The repeated application of the spin-raising operator turns $\ket{{\rm D}_{N,m}}$ into a fully polarised state: $\big( \hat S_+\big)^{\frac{N}{2}-m} \ket{{\rm D}_{N,m}} \propto \ket{{\rm D}_{N, \frac{N}{2}}}$. Since this state is fully polarised, $\hat \Pi \ket{{\rm D}_{N, \frac{N}{2}}}=0$.
By using the following expression for the spin-raising operator: $\hat S_+ = \sum_k \hat c_{k\uparrow}^\dagger \hat c_{k \downarrow}$, it is not difficult to show that $[\hat \Pi, \hat S_+]=0$ because $\hat n_{k,\uparrow} \hat n_{k,\downarrow} \hat c_{k\uparrow}^\dagger \hat c_{k \downarrow} = \hat c_{k\uparrow}^\dagger \hat c_{k \downarrow} \hat n_{k,\uparrow} \hat n_{k,\downarrow} = 0$. With this relation we can show that $\hat \Pi \ket{{\rm D}_{N,m}}=0$. From this we obtain that in general $\hat \Pi \ket{{\rm D}_N}=0$ and thus that also $\hat \Pi \ket{\Psi_{\rm D}}=0$. 
With similar reasonings it is possible to show also the other relations in~\eqref{eq:Dicke2} and this concludes the proof.
$\qed$

\section{Additional data for the band-insulator dynamics}\label{App:Additional:BI}
\begin{figure}[t]
 \includegraphics[width=\columnwidth]{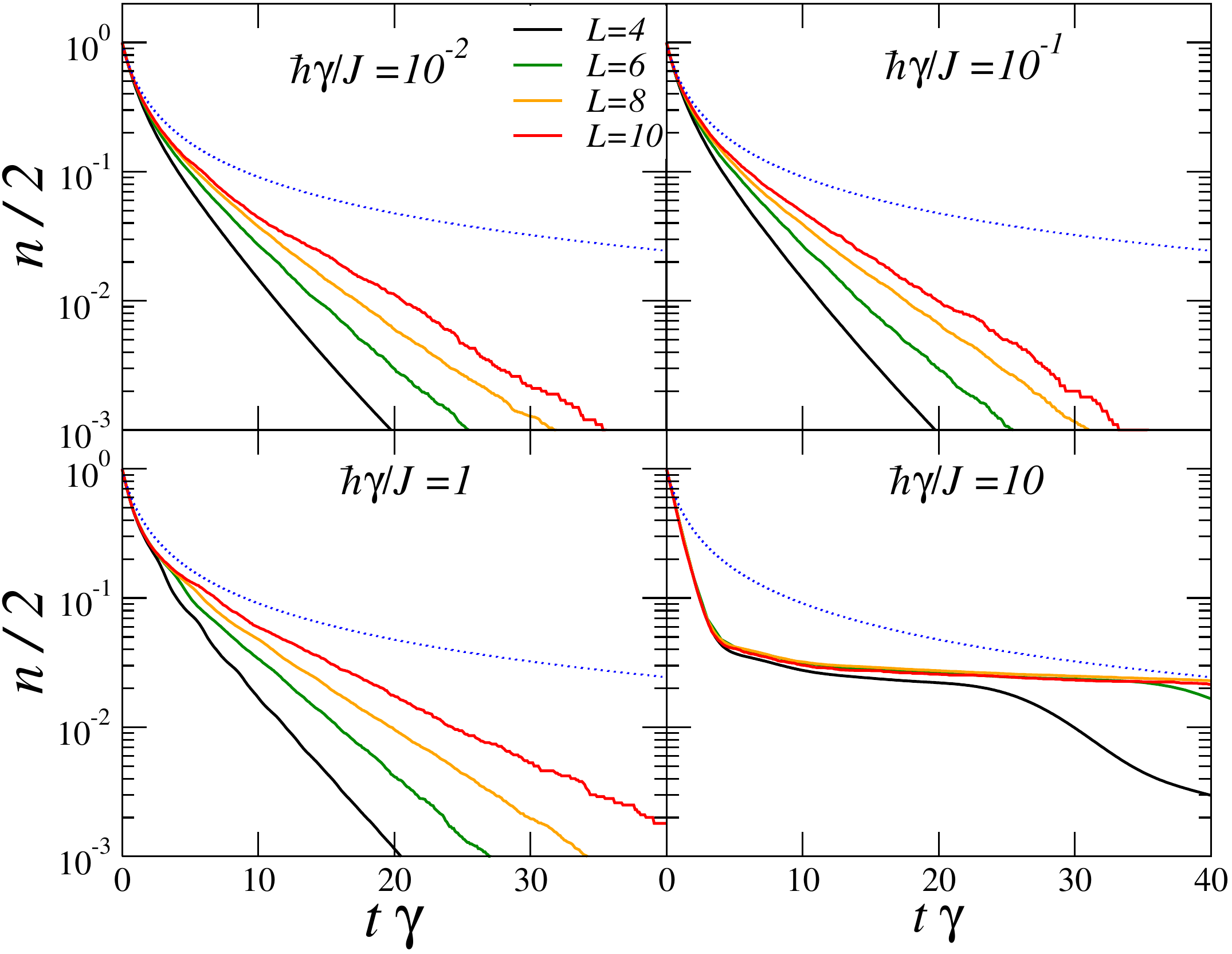}
 \caption{Dissipative dynamics of the normalised density a band insulator for $J/\gamma = 10^2$, $10$, $1$ and $0.1$. Different colors refer to different sizes, from $L=4$ to $L=10$. The blue dotted curve is the prediction for the thermodynamic limit in Eq.~\eqref{Eq:n:BI:TL} and highlights the importance of finite-size effects in our simulations. The appearance of a new behaviour in the strongly-dissipative limit is evident.}\label{Fig:BI}
\end{figure}

In this appendix we present some additional data from our quantum trajectory simulations of the band-insulator dissipative dynamics.
In Fig.~\ref{Fig:BI} we plot the same data as in Fig.~\ref{Fig:n:BI:LScaling} at fixed $J/\hbar \gamma$ and varying $L$. We observe the importance of finite-size effects and the absence of any collapse at the sizes that we could consider numerically. 

\begin{figure}[t]
  \includegraphics[width=\columnwidth]{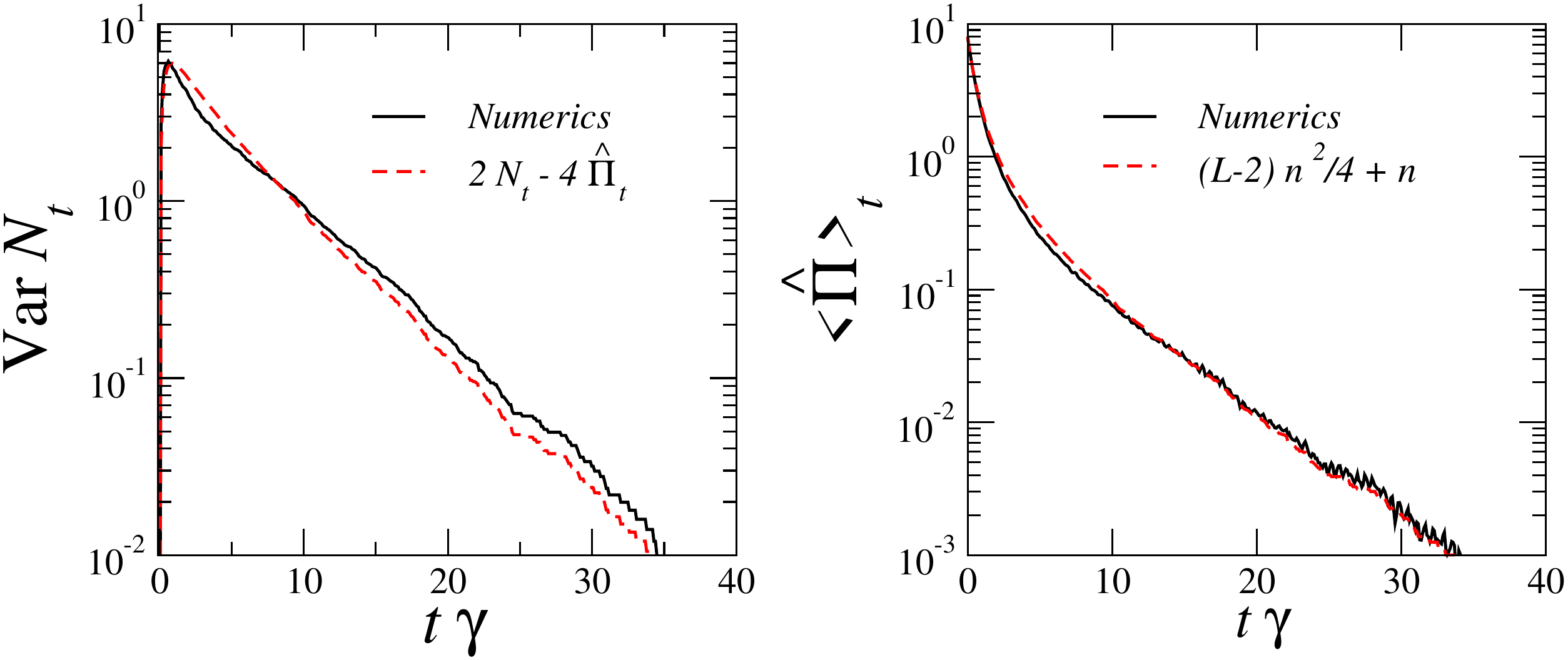}
  \caption{$\Var N_t$ (left panel) and $\langle \hat \Pi \rangle_t$ (right panel) for $L=8$ and $J/\hbar \gamma=10$. Black solid line: calculation of the two quantities using $2000$ quantum trajectories. Red dashed line: approximation using formulas~\eqref{Eq:BI:Var:Pi} and using the numerically-computed value for $N(t)$.}
  \label{Fig:BI:Pi:Var}
\end{figure}

In Fig.~\ref{Fig:BI:Pi:Var} we plot the numerically-computed $\Var N_t$ and $\langle \hat \Pi \rangle_t$ and compare them to the expressions proposed in Eqs.~\eqref{Eq:BI:Var:Pi}.

\section{Additional data on the calculation of the purity}
\label{App:Purity}
\begin{figure}[!t]
  \includegraphics[width=0.7\columnwidth]{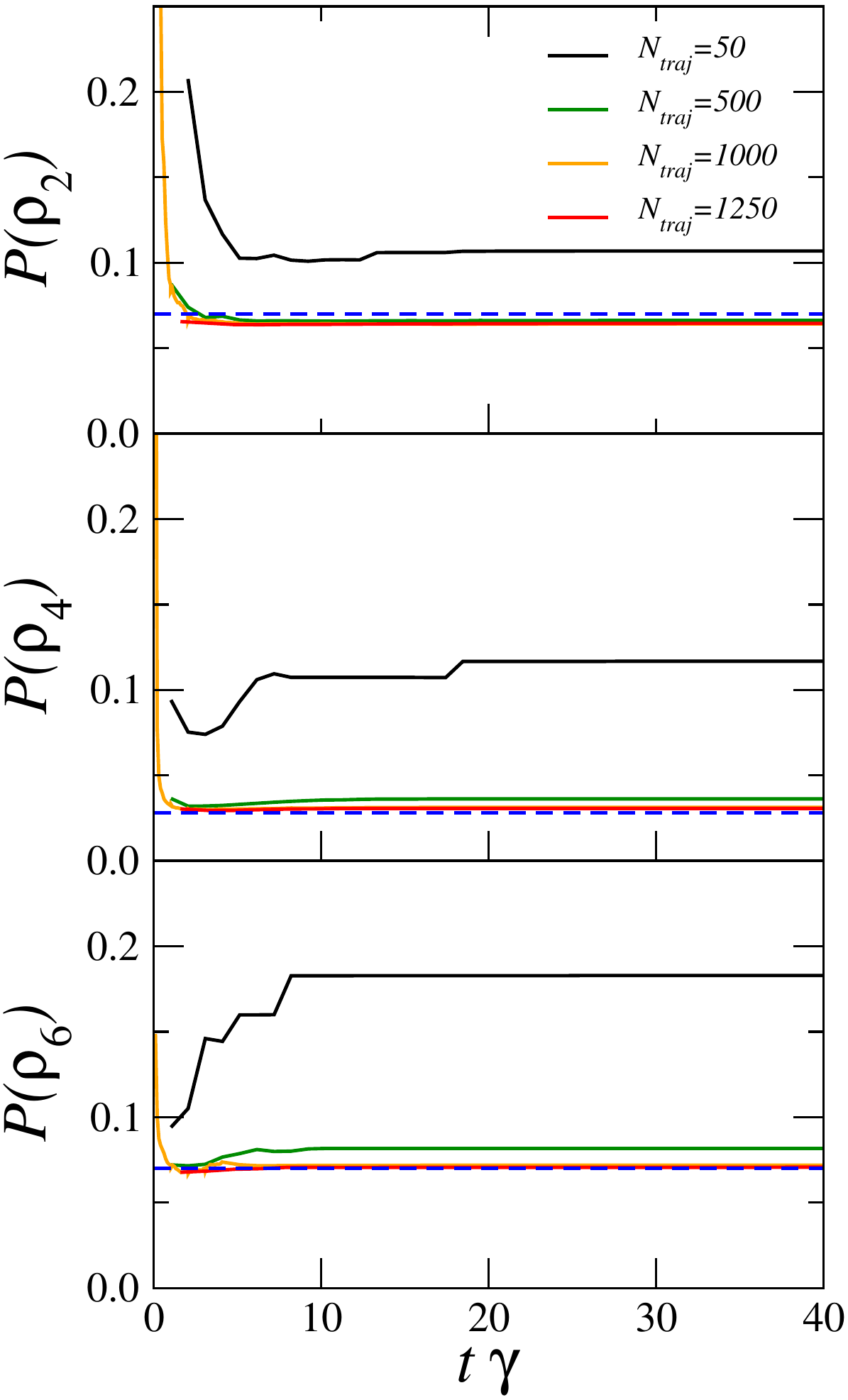} 
  \caption{Symmetry-resolved purity $\mathcal{P}\left(\hat\rho_n\right)$ for $n=2,4,6$ for different values of $N_{\rm traj}$. The dashed blue lines represent the theoretical prediction for the steady-state symmetry-resolved purity $\mathcal P(\rho_n) = 2/\binom{L}{n}$. Parameters are set as in Fig.~\ref{Fig:purity_mott_symres}.}
  \label{Fig:purity_mott_symres_convergence}
\end{figure}
We now comment on the convergence of the results presented in Sec.~\ref{Sec:Symm:Res:Pur} with respect to the number of stochastic trajectories. 
From Eq.~\eqref{rhotraj} it is clear that the degree of accuracy in the reconstruction of the density matrix depends on $N_{\rm traj}$. 
In particular, as $N_{\rm traj}$ is increased we can account for more statistically-independent realisations of the dynamics; intuitively, one expects $\mathcal{P}\left(\rho_n\right)$ to be a monotonically decreasing function of $N_{\rm traj}$.
This intuition is confirmed by the numerical data presented in Fig.~\ref{Fig:purity_mott_symres_convergence}.

Decomposing the symmetry-resolved purity in terms of diagonal and off-diagonal overlaps we get that
\begin{equation}
  \mathcal{P}(\rho_n) = \frac{1}{N_n} +\frac{1}{N_n^2} \sum_{\alpha\neq\beta}^{N_n}|\bra{\psi_{n,\alpha}} \psi_{n,\beta} \rangle|^2 .
\end{equation}
Since we are interested in the $N_n \to \infty$ limit, we observe that the limiting value of the purity can be obtained by only studying the second addend of the r.h.s.

We now make the assumption that $N_n\propto N_{\rm traj}$ and 
that the $\ket{\psi_{n,\alpha}}$ are randomly distributed in the dark subspace $\mathcal H_n^{\rm Dark}$, so that
the off-diagonal overlaps scale as $|\bra{\psi_{n,\alpha}} \psi_{n,\beta} \rangle|^2 \sim1/{{\rm dim} (\mathcal{H}_n^{\rm Dark})}$ for $\alpha\neq\beta$.
In the limit $N_n\to \infty$ we obtain: 
\begin{equation}
  \label{thformpur}
  \mathcal{P}(\rho_n) \simeq \frac{1}{{\rm dim}(\mathcal{H}_n^{\rm Dark})}.
\end{equation}

In our specific case, we can estimate the dimension of the dark subspace by counting the number of antisymmetric orbital wavefunction to be associated to the fully-symmetric spin part of the wavefunction, a Dicke state with $S=n/2$ and $S_z=0$. Thus, $\mathrm{dim} (\mathcal H^{\rm Dark}_n ) = \binom{L}{n}$.
Eq.~\eqref{thformpur} provides a very good estimation of the asymptotic value of the symmetry-resolved purity as shown in Fig.~\ref{Fig:purity_mott_symres_convergence} where also the convergence with $N_{\rm traj}$ is shown.

By numerical inspection, the asymptotic purity value is $\mathcal P(\rho_n) = 2/\binom{L}{n}$; we do not understand the reason for the $2$ factor appearing in the formula, that points at a lack of ergodicity and to the fact that the only half of the dark subspace is explored by the dynamics. We pose that this is due to the specific spin structure of the initial N\'eel state, and that a Mott insulator with randomly-oriented spins would explore the full dark state; we leave a more systematic study to the future. 

Finally, we stress that this numerical computation is quite heavy in terms of memory since it requires requires to allocate $N_{\rm traj}$ wave functions of the many-body system for many values of $t$, this limits our analysis to $N_{\rm traj}=1250$.

\bibliography{DissYbBos.bib}

\begin{thebibliography}{39}%
\makeatletter
\providecommand \@ifxundefined [1]{%
 \@ifx{#1\undefined}
}%
\providecommand \@ifnum [1]{%
 \ifnum #1\expandafter \@firstoftwo
 \else \expandafter \@secondoftwo
 \fi
}%
\providecommand \@ifx [1]{%
 \ifx #1\expandafter \@firstoftwo
 \else \expandafter \@secondoftwo
 \fi
}%
\providecommand \natexlab [1]{#1}%
\providecommand \enquote  [1]{``#1''}%
\providecommand \bibnamefont  [1]{#1}%
\providecommand \bibfnamefont [1]{#1}%
\providecommand \citenamefont [1]{#1}%
\providecommand \href@noop [0]{\@secondoftwo}%
\providecommand \href [0]{\begingroup \@sanitize@url \@href}%
\providecommand \@href[1]{\@@startlink{#1}\@@href}%
\providecommand \@@href[1]{\endgroup#1\@@endlink}%
\providecommand \@sanitize@url [0]{\catcode `\\12\catcode `\$12\catcode
  `\&12\catcode `\#12\catcode `\^12\catcode `\_12\catcode `\%12\relax}%
\providecommand \@@startlink[1]{}%
\providecommand \@@endlink[0]{}%
\providecommand \url  [0]{\begingroup\@sanitize@url \@url }%
\providecommand \@url [1]{\endgroup\@href {#1}{\urlprefix }}%
\providecommand \urlprefix  [0]{URL }%
\providecommand \Eprint [0]{\href }%
\providecommand \doibase [0]{https://doi.org/}%
\providecommand \selectlanguage [0]{\@gobble}%
\providecommand \bibinfo  [0]{\@secondoftwo}%
\providecommand \bibfield  [0]{\@secondoftwo}%
\providecommand \translation [1]{[#1]}%
\providecommand \BibitemOpen [0]{}%
\providecommand \bibitemStop [0]{}%
\providecommand \bibitemNoStop [0]{.\EOS\space}%
\providecommand \EOS [0]{\spacefactor3000\relax}%
\providecommand \BibitemShut  [1]{\csname bibitem#1\endcsname}%
\let\auto@bib@innerbib\@empty
\bibitem [{\citenamefont {Langen}\ \emph {et~al.}(2015)\citenamefont {Langen},
  \citenamefont {Geiger},\ and\ \citenamefont {Schmiedmayer}}]{LangenRev_2015}%
  \BibitemOpen
  \bibfield  {author} {\bibinfo {author} {\bibfnamefont {T.}~\bibnamefont
  {Langen}}, \bibinfo {author} {\bibfnamefont {R.}~\bibnamefont {Geiger}},\
  and\ \bibinfo {author} {\bibfnamefont {J.}~\bibnamefont {Schmiedmayer}},\
  }\bibfield  {title} {\bibinfo {title} {Ultracold atoms out of equilibrium},\
  }\href {https://doi.org/10.1146/annurev-conmatphys-031214-014548} {\bibfield
  {journal} {\bibinfo  {journal} {Annual Review of Condensed Matter Physics}\
  }\textbf {\bibinfo {volume} {6}},\ \bibinfo {pages} {201} (\bibinfo {year}
  {2015})}\BibitemShut {NoStop}%
\bibitem [{\citenamefont {Zurek}(2003)}]{Zurek_2003}%
  \BibitemOpen
  \bibfield  {author} {\bibinfo {author} {\bibfnamefont {W.~H.}\ \bibnamefont
  {Zurek}},\ }\bibfield  {title} {\bibinfo {title} {Decoherence, einselection,
  and the quantum origins of the classical},\ }\href
  {https://doi.org/10.1103/RevModPhys.75.715} {\bibfield  {journal} {\bibinfo
  {journal} {Rev. Mod. Phys.}\ }\textbf {\bibinfo {volume} {75}},\ \bibinfo
  {pages} {715} (\bibinfo {year} {2003})}\BibitemShut {NoStop}%
\bibitem [{\citenamefont {Laburthe~Tolra}\ \emph {et~al.}(2004)\citenamefont
  {Laburthe~Tolra}, \citenamefont {O'Hara}, \citenamefont {Huckans},
  \citenamefont {Phillips}, \citenamefont {Rolston},\ and\ \citenamefont
  {Porto}}]{Tolra_2004}%
  \BibitemOpen
  \bibfield  {author} {\bibinfo {author} {\bibfnamefont {B.}~\bibnamefont
  {Laburthe~Tolra}}, \bibinfo {author} {\bibfnamefont {K.~M.}\ \bibnamefont
  {O'Hara}}, \bibinfo {author} {\bibfnamefont {J.~H.}\ \bibnamefont {Huckans}},
  \bibinfo {author} {\bibfnamefont {W.~D.}\ \bibnamefont {Phillips}}, \bibinfo
  {author} {\bibfnamefont {S.~L.}\ \bibnamefont {Rolston}},\ and\ \bibinfo
  {author} {\bibfnamefont {J.~V.}\ \bibnamefont {Porto}},\ }\bibfield  {title}
  {\bibinfo {title} {Observation of reduced three-body recombination in a
  correlated 1d degenerate bose gas},\ }\href
  {https://doi.org/10.1103/PhysRevLett.92.190401} {\bibfield  {journal}
  {\bibinfo  {journal} {Phys. Rev. Lett.}\ }\textbf {\bibinfo {volume} {92}},\
  \bibinfo {pages} {190401} (\bibinfo {year} {2004})}\BibitemShut {NoStop}%
\bibitem [{\citenamefont {Kraemer}\ \emph {et~al.}(2006)\citenamefont
  {Kraemer}, \citenamefont {Mark}, \citenamefont {Waldburger}, \citenamefont
  {Danzl}, \citenamefont {Chin}, \citenamefont {Engeser}, \citenamefont
  {Lange}, \citenamefont {Pilch}, \citenamefont {Jaakkola}, \citenamefont
  {N\"agerl},\ and\ \citenamefont {Grimm}}]{Kraemer_2006}%
  \BibitemOpen
  \bibfield  {author} {\bibinfo {author} {\bibfnamefont {T.}~\bibnamefont
  {Kraemer}}, \bibinfo {author} {\bibfnamefont {M.}~\bibnamefont {Mark}},
  \bibinfo {author} {\bibfnamefont {P.}~\bibnamefont {Waldburger}}, \bibinfo
  {author} {\bibfnamefont {J.~G.}\ \bibnamefont {Danzl}}, \bibinfo {author}
  {\bibfnamefont {C.}~\bibnamefont {Chin}}, \bibinfo {author} {\bibfnamefont
  {B.}~\bibnamefont {Engeser}}, \bibinfo {author} {\bibfnamefont {A.~D.}\
  \bibnamefont {Lange}}, \bibinfo {author} {\bibfnamefont {K.}~\bibnamefont
  {Pilch}}, \bibinfo {author} {\bibfnamefont {A.}~\bibnamefont {Jaakkola}},
  \bibinfo {author} {\bibfnamefont {H.-C.}\ \bibnamefont {N\"agerl}},\ and\
  \bibinfo {author} {\bibfnamefont {R.}~\bibnamefont {Grimm}},\ }\bibfield
  {title} {\bibinfo {title} {Evidence for efimov quantum states in an ultracold
  gas of caesium atoms},\ }\href@noop {} {\bibfield  {journal} {\bibinfo
  {journal} {Nature}\ }\textbf {\bibinfo {volume} {440}},\ \bibinfo {pages}
  {315} (\bibinfo {year} {2006})}\BibitemShut {NoStop}%
\bibitem [{\citenamefont {Pollack}\ \emph {et~al.}(2009)\citenamefont
  {Pollack}, \citenamefont {Dries},\ and\ \citenamefont
  {Hulet}}]{Pollack_2009}%
  \BibitemOpen
  \bibfield  {author} {\bibinfo {author} {\bibfnamefont {S.~E.}\ \bibnamefont
  {Pollack}}, \bibinfo {author} {\bibfnamefont {D.}~\bibnamefont {Dries}},\
  and\ \bibinfo {author} {\bibfnamefont {R.~G.}\ \bibnamefont {Hulet}},\
  }\bibfield  {title} {\bibinfo {title} {Universality in three- and four-body
  bound states of ultracold atoms},\ }\href
  {https://doi.org/10.1126/science.1182840} {\bibfield  {journal} {\bibinfo
  {journal} {Science}\ }\textbf {\bibinfo {volume} {326}},\ \bibinfo {pages}
  {1683} (\bibinfo {year} {2009})}\BibitemShut {NoStop}%
\bibitem [{\citenamefont {Baur}\ and\ \citenamefont
  {Mueller}(2010)}]{Baur_2010}%
  \BibitemOpen
  \bibfield  {author} {\bibinfo {author} {\bibfnamefont {S.~K.}\ \bibnamefont
  {Baur}}\ and\ \bibinfo {author} {\bibfnamefont {E.~J.}\ \bibnamefont
  {Mueller}},\ }\bibfield  {title} {\bibinfo {title} {Two-body recombination in
  a quantum-mechanical lattice gas: Entropy generation and probing of
  short-range magnetic correlations},\ }\href
  {https://doi.org/10.1103/physreva.82.023626} {\bibfield  {journal} {\bibinfo
  {journal} {Physical Review A}\ }\textbf {\bibinfo {volume} {82}},\ \bibinfo
  {pages} {023626} (\bibinfo {year} {2010})}\BibitemShut {NoStop}%
\bibitem [{\citenamefont {Gri\ifmmode~\check{s}\else \v{s}\fi{}ins}\ \emph
  {et~al.}(2016)\citenamefont {Gri\ifmmode~\check{s}\else \v{s}\fi{}ins},
  \citenamefont {Rauer}, \citenamefont {Langen}, \citenamefont {Schmiedmayer},\
  and\ \citenamefont {Mazets}}]{Grisins_2016}%
  \BibitemOpen
  \bibfield  {author} {\bibinfo {author} {\bibfnamefont {P.}~\bibnamefont
  {Gri\ifmmode~\check{s}\else \v{s}\fi{}ins}}, \bibinfo {author} {\bibfnamefont
  {B.}~\bibnamefont {Rauer}}, \bibinfo {author} {\bibfnamefont
  {T.}~\bibnamefont {Langen}}, \bibinfo {author} {\bibfnamefont
  {J.}~\bibnamefont {Schmiedmayer}},\ and\ \bibinfo {author} {\bibfnamefont
  {I.~E.}\ \bibnamefont {Mazets}},\ }\bibfield  {title} {\bibinfo {title}
  {Degenerate bose gases with uniform loss},\ }\href
  {https://doi.org/10.1103/PhysRevA.93.033634} {\bibfield  {journal} {\bibinfo
  {journal} {Phys. Rev. A}\ }\textbf {\bibinfo {volume} {93}},\ \bibinfo
  {pages} {033634} (\bibinfo {year} {2016})}\BibitemShut {NoStop}%
\bibitem [{\citenamefont {Bouchoule}\ \emph {et~al.}(2018)\citenamefont
  {Bouchoule}, \citenamefont {Schemmer},\ and\ \citenamefont
  {Henkel}}]{Bouchoule_2018}%
  \BibitemOpen
  \bibfield  {author} {\bibinfo {author} {\bibfnamefont {I.}~\bibnamefont
  {Bouchoule}}, \bibinfo {author} {\bibfnamefont {M.}~\bibnamefont
  {Schemmer}},\ and\ \bibinfo {author} {\bibfnamefont {C.}~\bibnamefont
  {Henkel}},\ }\bibfield  {title} {\bibinfo {title} {Cooling phonon modes of a
  bose condensate with uniform few body losses},\ }\href
  {https://doi.org/10.21468/scipostphys.5.5.043} {\bibfield  {journal}
  {\bibinfo  {journal} {SciPost Physics}\ }\textbf {\bibinfo {volume} {5}},\
  \bibinfo {pages} {043} (\bibinfo {year} {2018})}\BibitemShut {NoStop}%
\bibitem [{\citenamefont {Schemmer}\ and\ \citenamefont
  {Bouchoule}(2018)}]{Schemmer_2018}%
  \BibitemOpen
  \bibfield  {author} {\bibinfo {author} {\bibfnamefont {M.}~\bibnamefont
  {Schemmer}}\ and\ \bibinfo {author} {\bibfnamefont {I.}~\bibnamefont
  {Bouchoule}},\ }\bibfield  {title} {\bibinfo {title} {Cooling a bose gas by
  three-body losses},\ }\href {https://doi.org/10.1103/PhysRevLett.121.200401}
  {\bibfield  {journal} {\bibinfo  {journal} {Phys. Rev. Lett.}\ }\textbf
  {\bibinfo {volume} {121}},\ \bibinfo {pages} {200401} (\bibinfo {year}
  {2018})}\BibitemShut {NoStop}%
\bibitem [{\citenamefont {Dogra}\ \emph {et~al.}(2019)\citenamefont {Dogra},
  \citenamefont {Glidden}, \citenamefont {Hilker}, \citenamefont {Eigen},
  \citenamefont {Cornell}, \citenamefont {Smith},\ and\ \citenamefont
  {Hadzibabic}}]{Dogra_2019}%
  \BibitemOpen
  \bibfield  {author} {\bibinfo {author} {\bibfnamefont {L.~H.}\ \bibnamefont
  {Dogra}}, \bibinfo {author} {\bibfnamefont {J.~A.~P.}\ \bibnamefont
  {Glidden}}, \bibinfo {author} {\bibfnamefont {T.~A.}\ \bibnamefont {Hilker}},
  \bibinfo {author} {\bibfnamefont {C.}~\bibnamefont {Eigen}}, \bibinfo
  {author} {\bibfnamefont {E.~A.}\ \bibnamefont {Cornell}}, \bibinfo {author}
  {\bibfnamefont {R.~P.}\ \bibnamefont {Smith}},\ and\ \bibinfo {author}
  {\bibfnamefont {Z.}~\bibnamefont {Hadzibabic}},\ }\bibfield  {title}
  {\bibinfo {title} {Can three-body recombination purify a quantum gas?},\
  }\href {https://doi.org/10.1103/PhysRevLett.123.020405} {\bibfield  {journal}
  {\bibinfo  {journal} {Phys. Rev. Lett.}\ }\textbf {\bibinfo {volume} {123}},\
  \bibinfo {pages} {020405} (\bibinfo {year} {2019})}\BibitemShut {NoStop}%
\bibitem [{\citenamefont {Syassen}\ \emph {et~al.}(2008)\citenamefont
  {Syassen}, \citenamefont {Bauer}, \citenamefont {Lettner}, \citenamefont
  {Volz}, \citenamefont {Dietze}, \citenamefont {Garc{\'\i}a-Ripoll},
  \citenamefont {Cirac}, \citenamefont {Rempe},\ and\ \citenamefont
  {D{\"u}rr}}]{Syassen_2008}%
  \BibitemOpen
  \bibfield  {author} {\bibinfo {author} {\bibfnamefont {N.}~\bibnamefont
  {Syassen}}, \bibinfo {author} {\bibfnamefont {D.~M.}\ \bibnamefont {Bauer}},
  \bibinfo {author} {\bibfnamefont {M.}~\bibnamefont {Lettner}}, \bibinfo
  {author} {\bibfnamefont {T.}~\bibnamefont {Volz}}, \bibinfo {author}
  {\bibfnamefont {D.}~\bibnamefont {Dietze}}, \bibinfo {author} {\bibfnamefont
  {J.~J.}\ \bibnamefont {Garc{\'\i}a-Ripoll}}, \bibinfo {author} {\bibfnamefont
  {J.~I.}\ \bibnamefont {Cirac}}, \bibinfo {author} {\bibfnamefont
  {G.}~\bibnamefont {Rempe}},\ and\ \bibinfo {author} {\bibfnamefont
  {S.}~\bibnamefont {D{\"u}rr}},\ }\bibfield  {title} {\bibinfo {title} {Strong
  dissipation inhibits losses and induces correlations in cold molecular
  gases},\ }\href {https://doi.org/10.1126/science.1155309} {\bibfield
  {journal} {\bibinfo  {journal} {Science}\ }\textbf {\bibinfo {volume}
  {320}},\ \bibinfo {pages} {1329} (\bibinfo {year} {2008})}\BibitemShut
  {NoStop}%
\bibitem [{\citenamefont {Garc{\'{\i}}a-Ripoll}\ \emph
  {et~al.}(2009)\citenamefont {Garc{\'{\i}}a-Ripoll}, \citenamefont {D\"urr},
  \citenamefont {Syassen}, \citenamefont {Bauer}, \citenamefont {Lettner},
  \citenamefont {Rempe},\ and\ \citenamefont {Cirac}}]{GarciaRipoll_2009}%
  \BibitemOpen
  \bibfield  {author} {\bibinfo {author} {\bibfnamefont {J.~J.}\ \bibnamefont
  {Garc{\'{\i}}a-Ripoll}}, \bibinfo {author} {\bibfnamefont {S.}~\bibnamefont
  {D\"urr}}, \bibinfo {author} {\bibfnamefont {N.}~\bibnamefont {Syassen}},
  \bibinfo {author} {\bibfnamefont {D.~M.}\ \bibnamefont {Bauer}}, \bibinfo
  {author} {\bibfnamefont {M.}~\bibnamefont {Lettner}}, \bibinfo {author}
  {\bibfnamefont {G.}~\bibnamefont {Rempe}},\ and\ \bibinfo {author}
  {\bibfnamefont {J.~I.}\ \bibnamefont {Cirac}},\ }\bibfield  {title} {\bibinfo
  {title} {Dissipation-induced hard-core boson gas in an optical lattice},\
  }\href {https://doi.org/10.1088/1367-2630/11/1/013053} {\bibfield  {journal}
  {\bibinfo  {journal} {New J. Phys.}\ }\textbf {\bibinfo {volume} {11}},\
  \bibinfo {pages} {013053} (\bibinfo {year} {2009})}\BibitemShut {NoStop}%
\bibitem [{\citenamefont {Kantian}\ \emph {et~al.}(2009)\citenamefont
  {Kantian}, \citenamefont {Dalmonte}, \citenamefont {Diehl}, \citenamefont
  {Hofstetter}, \citenamefont {Zoller},\ and\ \citenamefont
  {Daley}}]{Kantian_2009}%
  \BibitemOpen
  \bibfield  {author} {\bibinfo {author} {\bibfnamefont {A.}~\bibnamefont
  {Kantian}}, \bibinfo {author} {\bibfnamefont {M.}~\bibnamefont {Dalmonte}},
  \bibinfo {author} {\bibfnamefont {S.}~\bibnamefont {Diehl}}, \bibinfo
  {author} {\bibfnamefont {W.}~\bibnamefont {Hofstetter}}, \bibinfo {author}
  {\bibfnamefont {P.}~\bibnamefont {Zoller}},\ and\ \bibinfo {author}
  {\bibfnamefont {A.~J.}\ \bibnamefont {Daley}},\ }\bibfield  {title} {\bibinfo
  {title} {Atomic color superfluid via three-body loss},\ }\href
  {https://doi.org/10.1103/PhysRevLett.103.240401} {\bibfield  {journal}
  {\bibinfo  {journal} {Phys. Rev. Lett.}\ }\textbf {\bibinfo {volume} {103}},\
  \bibinfo {pages} {240401} (\bibinfo {year} {2009})}\BibitemShut {NoStop}%
\bibitem [{\citenamefont {Daley}\ \emph {et~al.}(2009)\citenamefont {Daley},
  \citenamefont {Taylor}, \citenamefont {Diehl}, \citenamefont {Baranov},\ and\
  \citenamefont {Zoller}}]{Daley_2009}%
  \BibitemOpen
  \bibfield  {author} {\bibinfo {author} {\bibfnamefont {A.~J.}\ \bibnamefont
  {Daley}}, \bibinfo {author} {\bibfnamefont {J.~M.}\ \bibnamefont {Taylor}},
  \bibinfo {author} {\bibfnamefont {S.}~\bibnamefont {Diehl}}, \bibinfo
  {author} {\bibfnamefont {M.}~\bibnamefont {Baranov}},\ and\ \bibinfo {author}
  {\bibfnamefont {P.}~\bibnamefont {Zoller}},\ }\bibfield  {title} {\bibinfo
  {title} {Atomic three-body loss as a dynamical three-body interaction},\
  }\href {https://doi.org/10.1103/PhysRevLett.102.040402} {\bibfield  {journal}
  {\bibinfo  {journal} {Phys. Rev. Lett.}\ }\textbf {\bibinfo {volume} {102}},\
  \bibinfo {pages} {040402} (\bibinfo {year} {2009})}\BibitemShut {NoStop}%
\bibitem [{\citenamefont {Roncaglia}\ \emph {et~al.}(2010)\citenamefont
  {Roncaglia}, \citenamefont {Rizzi},\ and\ \citenamefont
  {Cirac}}]{Roncaglia_2010}%
  \BibitemOpen
  \bibfield  {author} {\bibinfo {author} {\bibfnamefont {M.}~\bibnamefont
  {Roncaglia}}, \bibinfo {author} {\bibfnamefont {M.}~\bibnamefont {Rizzi}},\
  and\ \bibinfo {author} {\bibfnamefont {J.~I.}\ \bibnamefont {Cirac}},\
  }\bibfield  {title} {\bibinfo {title} {Pfaffian state generation by strong
  three-body dissipation},\ }\href
  {https://doi.org/10.1103/PhysRevLett.104.096803} {\bibfield  {journal}
  {\bibinfo  {journal} {Phys. Rev. Lett.}\ }\textbf {\bibinfo {volume} {104}},\
  \bibinfo {pages} {096803} (\bibinfo {year} {2010})}\BibitemShut {NoStop}%
\bibitem [{\citenamefont {Beige}\ \emph {et~al.}(2000)\citenamefont {Beige},
  \citenamefont {Braun}, \citenamefont {Tregenna},\ and\ \citenamefont
  {Knight}}]{Almut_2000}%
  \BibitemOpen
  \bibfield  {author} {\bibinfo {author} {\bibfnamefont {A.}~\bibnamefont
  {Beige}}, \bibinfo {author} {\bibfnamefont {D.}~\bibnamefont {Braun}},
  \bibinfo {author} {\bibfnamefont {B.}~\bibnamefont {Tregenna}},\ and\
  \bibinfo {author} {\bibfnamefont {P.~L.}\ \bibnamefont {Knight}},\ }\bibfield
   {title} {\bibinfo {title} {Quantum computing using dissipation to remain in
  a decoherence-free subspace},\ }\href
  {https://doi.org/10.1103/PhysRevLett.85.1762} {\bibfield  {journal} {\bibinfo
   {journal} {Phys. Rev. Lett.}\ }\textbf {\bibinfo {volume} {85}},\ \bibinfo
  {pages} {1762} (\bibinfo {year} {2000})}\BibitemShut {NoStop}%
\bibitem [{\citenamefont {Diehl}\ \emph {et~al.}(2008)\citenamefont {Diehl},
  \citenamefont {Micheli}, \citenamefont {Kantian}, \citenamefont {Kraus},
  \citenamefont {B\"uchler},\ and\ \citenamefont {Zoller}}]{Diehl_2008}%
  \BibitemOpen
  \bibfield  {author} {\bibinfo {author} {\bibfnamefont {S.}~\bibnamefont
  {Diehl}}, \bibinfo {author} {\bibfnamefont {A.}~\bibnamefont {Micheli}},
  \bibinfo {author} {\bibfnamefont {A.}~\bibnamefont {Kantian}}, \bibinfo
  {author} {\bibfnamefont {B.}~\bibnamefont {Kraus}}, \bibinfo {author}
  {\bibfnamefont {H.-P.}\ \bibnamefont {B\"uchler}},\ and\ \bibinfo {author}
  {\bibfnamefont {P.}~\bibnamefont {Zoller}},\ }\bibfield  {title} {\bibinfo
  {title} {Quantum states and phases in driven open quantum systems with cold
  atoms},\ }\href {https://doi.org/10.1038/nphys1073} {\bibfield  {journal}
  {\bibinfo  {journal} {Nat. Phys.}\ }\textbf {\bibinfo {volume} {4}},\
  \bibinfo {pages} {878} (\bibinfo {year} {2008})}\BibitemShut {NoStop}%
\bibitem [{\citenamefont {Verstraete}\ \emph {et~al.}(2009)\citenamefont
  {Verstraete}, \citenamefont {Wolf},\ and\ \citenamefont
  {Cirac}}]{Verstraete_2009}%
  \BibitemOpen
  \bibfield  {author} {\bibinfo {author} {\bibfnamefont {F.}~\bibnamefont
  {Verstraete}}, \bibinfo {author} {\bibfnamefont {M.~M.}\ \bibnamefont
  {Wolf}},\ and\ \bibinfo {author} {\bibfnamefont {J.~I.}\ \bibnamefont
  {Cirac}},\ }\bibfield  {title} {\bibinfo {title} {Quantum computation,
  quantum state engineering, and quantum phase transitions driven by
  dissipation},\ }\href {https://doi.org/10.1038/nphys1342} {\bibfield
  {journal} {\bibinfo  {journal} {Nat. Phys.}\ }\textbf {\bibinfo {volume}
  {5}},\ \bibinfo {pages} {633} (\bibinfo {year} {2009})}\BibitemShut {NoStop}%
\bibitem [{\citenamefont {Diehl}\ \emph {et~al.}(2011)\citenamefont {Diehl},
  \citenamefont {Rico}, \citenamefont {Baranov},\ and\ \citenamefont
  {Zoller}}]{Diehl_2011}%
  \BibitemOpen
  \bibfield  {author} {\bibinfo {author} {\bibfnamefont {S.}~\bibnamefont
  {Diehl}}, \bibinfo {author} {\bibfnamefont {E.}~\bibnamefont {Rico}},
  \bibinfo {author} {\bibfnamefont {M.~A.}\ \bibnamefont {Baranov}},\ and\
  \bibinfo {author} {\bibfnamefont {P.}~\bibnamefont {Zoller}},\ }\bibfield
  {title} {\bibinfo {title} {Topology by dissipation in atomic quantum wires},\
  }\href {https://doi.org/10.1038/nphys2106} {\bibfield  {journal} {\bibinfo
  {journal} {Nature Physics}\ }\textbf {\bibinfo {volume} {7}},\ \bibinfo
  {pages} {971} (\bibinfo {year} {2011})}\BibitemShut {NoStop}%
\bibitem [{\citenamefont {Iemini}\ \emph {et~al.}(2016)\citenamefont {Iemini},
  \citenamefont {Rossini}, \citenamefont {Fazio}, \citenamefont {Diehl},\ and\
  \citenamefont {Mazza}}]{Iemini_2016}%
  \BibitemOpen
  \bibfield  {author} {\bibinfo {author} {\bibfnamefont {F.}~\bibnamefont
  {Iemini}}, \bibinfo {author} {\bibfnamefont {D.}~\bibnamefont {Rossini}},
  \bibinfo {author} {\bibfnamefont {R.}~\bibnamefont {Fazio}}, \bibinfo
  {author} {\bibfnamefont {S.}~\bibnamefont {Diehl}},\ and\ \bibinfo {author}
  {\bibfnamefont {L.}~\bibnamefont {Mazza}},\ }\bibfield  {title} {\bibinfo
  {title} {Dissipative topological superconductors in number-conserving
  systems},\ }\href {https://doi.org/10.1103/PhysRevB.93.115113} {\bibfield
  {journal} {\bibinfo  {journal} {Phys. Rev. B}\ }\textbf {\bibinfo {volume}
  {93}},\ \bibinfo {pages} {115113} (\bibinfo {year} {2016})}\BibitemShut
  {NoStop}%
\bibitem [{\citenamefont {Kordas}\ \emph {et~al.}(2015)\citenamefont {Kordas},
  \citenamefont {Witthaut}, \citenamefont {Buonsante}, \citenamefont {Vezzani},
  \citenamefont {Burioni}, \citenamefont {Karanikas},\ and\ \citenamefont
  {Wimberger}}]{Kordas_2015}%
  \BibitemOpen
  \bibfield  {author} {\bibinfo {author} {\bibfnamefont {G.}~\bibnamefont
  {Kordas}}, \bibinfo {author} {\bibfnamefont {D.}~\bibnamefont {Witthaut}},
  \bibinfo {author} {\bibfnamefont {P.}~\bibnamefont {Buonsante}}, \bibinfo
  {author} {\bibfnamefont {A.}~\bibnamefont {Vezzani}}, \bibinfo {author}
  {\bibfnamefont {R.}~\bibnamefont {Burioni}}, \bibinfo {author} {\bibfnamefont
  {A.~I.}\ \bibnamefont {Karanikas}},\ and\ \bibinfo {author} {\bibfnamefont
  {S.}~\bibnamefont {Wimberger}},\ }\bibfield  {title} {\bibinfo {title} {The
  dissipative bose-hubbard model},\ }\href
  {https://doi.org/10.1140/epjst/e2015-02528-2} {\bibfield  {journal} {\bibinfo
   {journal} {The European Physical Journal Special Topics}\ }\textbf {\bibinfo
  {volume} {224}},\ \bibinfo {pages} {2127} (\bibinfo {year}
  {2015})}\BibitemShut {NoStop}%
\bibitem [{\citenamefont {Goto}\ and\ \citenamefont
  {Danshita}(2020)}]{Goto_2020}%
  \BibitemOpen
  \bibfield  {author} {\bibinfo {author} {\bibfnamefont {S.}~\bibnamefont
  {Goto}}\ and\ \bibinfo {author} {\bibfnamefont {I.}~\bibnamefont
  {Danshita}},\ }\bibfield  {title} {\bibinfo {title} {Measurement-induced
  transitions of the entanglement scaling law in ultracold gases with
  controllable dissipation},\ }\href
  {https://doi.org/10.1103/PhysRevA.102.033316} {\bibfield  {journal} {\bibinfo
   {journal} {Phys. Rev. A}\ }\textbf {\bibinfo {volume} {102}},\ \bibinfo
  {pages} {033316} (\bibinfo {year} {2020})}\BibitemShut {NoStop}%
\bibitem [{\citenamefont {Bouchoule}\ \emph {et~al.}(2020)\citenamefont
  {Bouchoule}, \citenamefont {Doyon},\ and\ \citenamefont
  {Dubail}}]{Bouchoule_2020b}%
  \BibitemOpen
  \bibfield  {author} {\bibinfo {author} {\bibfnamefont {I.}~\bibnamefont
  {Bouchoule}}, \bibinfo {author} {\bibfnamefont {B.}~\bibnamefont {Doyon}},\
  and\ \bibinfo {author} {\bibfnamefont {J.}~\bibnamefont {Dubail}},\
  }\bibfield  {title} {\bibinfo {title} {{The effect of atom losses on the
  distribution of rapidities in the one-dimensional Bose gas}},\ }\href
  {https://doi.org/10.21468/SciPostPhys.9.4.044} {\bibfield  {journal}
  {\bibinfo  {journal} {SciPost Phys.}\ }\textbf {\bibinfo {volume} {9}},\
  \bibinfo {pages} {44} (\bibinfo {year} {2020})}\BibitemShut {NoStop}%
\bibitem [{\citenamefont {Rossini}\ \emph {et~al.}(2020)\citenamefont
  {Rossini}, \citenamefont {Ghermaoui}, \citenamefont {Aguilera}, \citenamefont
  {Vatré}, \citenamefont {Bouganne}, \citenamefont {Beugnon}, \citenamefont
  {Gerbier},\ and\ \citenamefont {Mazza}}]{Rossini_2020}%
  \BibitemOpen
  \bibfield  {author} {\bibinfo {author} {\bibfnamefont {D.}~\bibnamefont
  {Rossini}}, \bibinfo {author} {\bibfnamefont {A.}~\bibnamefont {Ghermaoui}},
  \bibinfo {author} {\bibfnamefont {M.~B.}\ \bibnamefont {Aguilera}}, \bibinfo
  {author} {\bibfnamefont {R.}~\bibnamefont {Vatré}}, \bibinfo {author}
  {\bibfnamefont {R.}~\bibnamefont {Bouganne}}, \bibinfo {author}
  {\bibfnamefont {J.}~\bibnamefont {Beugnon}}, \bibinfo {author} {\bibfnamefont
  {F.}~\bibnamefont {Gerbier}},\ and\ \bibinfo {author} {\bibfnamefont
  {L.}~\bibnamefont {Mazza}},\ }\bibfield  {title} {\bibinfo {title} {Strong
  correlations in lossy one-dimensional quantum gases: from the quantum zeno
  effect to the generalized gibbs ensemble},\ }\href
  {https://doi.org/https://arxiv.org/abs/2011.04318} {\bibfield  {journal}
  {\bibinfo  {journal} {arXiv:2011.04318}\ ,\ } (\bibinfo {year}
  {2020})}\BibitemShut {NoStop}%
\bibitem [{\citenamefont {Bouchoule}\ and\ \citenamefont
  {Dubail}(2021)}]{Bouchoule_2021}%
  \BibitemOpen
  \bibfield  {author} {\bibinfo {author} {\bibfnamefont {I.}~\bibnamefont
  {Bouchoule}}\ and\ \bibinfo {author} {\bibfnamefont {J.}~\bibnamefont
  {Dubail}},\ }\bibfield  {title} {\bibinfo {title} {Breakdown of tan's
  relation in lossy one-dimensional bose gases},\ }\href@noop {} {\bibfield
  {journal} {\bibinfo  {journal} {arXiv:2011.13250}\ ,\ } (\bibinfo {year}
  {2021})}\BibitemShut {NoStop}%
\bibitem [{\citenamefont {Nakagawa}\ \emph {et~al.}(2021)\citenamefont
  {Nakagawa}, \citenamefont {Kawakami},\ and\ \citenamefont
  {Ueda}}]{Nakagawa_2021}%
  \BibitemOpen
  \bibfield  {author} {\bibinfo {author} {\bibfnamefont {M.}~\bibnamefont
  {Nakagawa}}, \bibinfo {author} {\bibfnamefont {N.}~\bibnamefont {Kawakami}},\
  and\ \bibinfo {author} {\bibfnamefont {M.}~\bibnamefont {Ueda}},\ }\bibfield
  {title} {\bibinfo {title} {Exact liouvillian spectrum of a one-dimensional
  dissipative hubbard model},\ }\href
  {https://doi.org/10.1103/PhysRevLett.126.110404} {\bibfield  {journal}
  {\bibinfo  {journal} {Phys. Rev. Lett.}\ }\textbf {\bibinfo {volume} {126}},\
  \bibinfo {pages} {110404} (\bibinfo {year} {2021})}\BibitemShut {NoStop}%
\bibitem [{\citenamefont {Foss-Feig}\ \emph {et~al.}(2012)\citenamefont
  {Foss-Feig}, \citenamefont {Daley}, \citenamefont {Thompson},\ and\
  \citenamefont {Rey}}]{FossFeig_2012}%
  \BibitemOpen
  \bibfield  {author} {\bibinfo {author} {\bibfnamefont {M.}~\bibnamefont
  {Foss-Feig}}, \bibinfo {author} {\bibfnamefont {A.~J.}\ \bibnamefont
  {Daley}}, \bibinfo {author} {\bibfnamefont {J.~K.}\ \bibnamefont
  {Thompson}},\ and\ \bibinfo {author} {\bibfnamefont {A.~M.}\ \bibnamefont
  {Rey}},\ }\bibfield  {title} {\bibinfo {title} {Steady-state many-body
  entanglement of hot reactive fermions},\ }\href
  {https://doi.org/10.1103/PhysRevLett.109.230501} {\bibfield  {journal}
  {\bibinfo  {journal} {Phys. Rev. Lett.}\ }\textbf {\bibinfo {volume} {109}},\
  \bibinfo {pages} {230501} (\bibinfo {year} {2012})}\BibitemShut {NoStop}%
\bibitem [{\citenamefont {Yan}\ \emph {et~al.}(2013)\citenamefont {Yan},
  \citenamefont {Moses}, \citenamefont {Gadway}, \citenamefont {Covey},
  \citenamefont {Hazzard}, \citenamefont {Rey}, \citenamefont {Jin},\ and\
  \citenamefont {Ye}}]{Yan_2013}%
  \BibitemOpen
  \bibfield  {author} {\bibinfo {author} {\bibfnamefont {B.}~\bibnamefont
  {Yan}}, \bibinfo {author} {\bibfnamefont {S.~A.}\ \bibnamefont {Moses}},
  \bibinfo {author} {\bibfnamefont {B.}~\bibnamefont {Gadway}}, \bibinfo
  {author} {\bibfnamefont {J.}~\bibnamefont {Covey}}, \bibinfo {author}
  {\bibfnamefont {K.~R.~A.}\ \bibnamefont {Hazzard}}, \bibinfo {author}
  {\bibfnamefont {A.~M.}\ \bibnamefont {Rey}}, \bibinfo {author} {\bibfnamefont
  {D.~S.}\ \bibnamefont {Jin}},\ and\ \bibinfo {author} {\bibfnamefont
  {J.}~\bibnamefont {Ye}},\ }\href {https://doi.org/10.1038/nature12483}
  {\bibfield  {journal} {\bibinfo  {journal} {Nature}\ }\textbf {\bibinfo
  {volume} {501}},\ \bibinfo {pages} {521} (\bibinfo {year}
  {2013})}\BibitemShut {NoStop}%
\bibitem [{\citenamefont {Zhu}\ \emph {et~al.}(2014)\citenamefont {Zhu},
  \citenamefont {Gadway}, \citenamefont {Foss-Feig}, \citenamefont
  {Schachenmayer}, \citenamefont {Wall}, \citenamefont {Hazzard}, \citenamefont
  {Yan}, \citenamefont {Moses}, \citenamefont {Covey}, \citenamefont {Jin},
  \citenamefont {Ye}, \citenamefont {Holland},\ and\ \citenamefont
  {Rey}}]{Zhu_2014}%
  \BibitemOpen
  \bibfield  {author} {\bibinfo {author} {\bibfnamefont {B.}~\bibnamefont
  {Zhu}}, \bibinfo {author} {\bibfnamefont {B.}~\bibnamefont {Gadway}},
  \bibinfo {author} {\bibfnamefont {M.}~\bibnamefont {Foss-Feig}}, \bibinfo
  {author} {\bibfnamefont {J.}~\bibnamefont {Schachenmayer}}, \bibinfo {author}
  {\bibfnamefont {M.~L.}\ \bibnamefont {Wall}}, \bibinfo {author}
  {\bibfnamefont {K.~R.~A.}\ \bibnamefont {Hazzard}}, \bibinfo {author}
  {\bibfnamefont {B.}~\bibnamefont {Yan}}, \bibinfo {author} {\bibfnamefont
  {S.~A.}\ \bibnamefont {Moses}}, \bibinfo {author} {\bibfnamefont {J.~P.}\
  \bibnamefont {Covey}}, \bibinfo {author} {\bibfnamefont {D.~S.}\ \bibnamefont
  {Jin}}, \bibinfo {author} {\bibfnamefont {J.}~\bibnamefont {Ye}}, \bibinfo
  {author} {\bibfnamefont {M.}~\bibnamefont {Holland}},\ and\ \bibinfo {author}
  {\bibfnamefont {A.~M.}\ \bibnamefont {Rey}},\ }\bibfield  {title} {\bibinfo
  {title} {Suppressing the loss of ultracold molecules via the continuous
  quantum zeno effect},\ }\href
  {https://doi.org/10.1103/PhysRevLett.112.070404} {\bibfield  {journal}
  {\bibinfo  {journal} {Phys. Rev. Lett.}\ }\textbf {\bibinfo {volume} {112}},\
  \bibinfo {pages} {070404} (\bibinfo {year} {2014})}\BibitemShut {NoStop}%
\bibitem [{\citenamefont {Sponselee}\ \emph {et~al.}(2019)\citenamefont
  {Sponselee}, \citenamefont {Freystatzky}, \citenamefont {Abeln},
  \citenamefont {Diem}, \citenamefont {Hundt}, \citenamefont {Kochanke},
  \citenamefont {Ponath}, \citenamefont {Santra}, \citenamefont {Mathey},
  \citenamefont {Sengstock},\ and\ \citenamefont {Becker}}]{Sponselee_2018}%
  \BibitemOpen
  \bibfield  {author} {\bibinfo {author} {\bibfnamefont {K.}~\bibnamefont
  {Sponselee}}, \bibinfo {author} {\bibfnamefont {L.}~\bibnamefont
  {Freystatzky}}, \bibinfo {author} {\bibfnamefont {B.}~\bibnamefont {Abeln}},
  \bibinfo {author} {\bibfnamefont {M.}~\bibnamefont {Diem}}, \bibinfo {author}
  {\bibfnamefont {B.}~\bibnamefont {Hundt}}, \bibinfo {author} {\bibfnamefont
  {A.}~\bibnamefont {Kochanke}}, \bibinfo {author} {\bibfnamefont
  {T.}~\bibnamefont {Ponath}}, \bibinfo {author} {\bibfnamefont
  {B.}~\bibnamefont {Santra}}, \bibinfo {author} {\bibfnamefont
  {L.}~\bibnamefont {Mathey}}, \bibinfo {author} {\bibfnamefont
  {K.}~\bibnamefont {Sengstock}},\ and\ \bibinfo {author} {\bibfnamefont
  {C.}~\bibnamefont {Becker}},\ }\bibfield  {title} {\bibinfo {title} {Dynamics
  of ultracold quantum gases in the dissipative fermi-hubbard model},\ }\href
  {https://doi.org/10.1088/2058-9565/aadccd} {\bibfield  {journal} {\bibinfo
  {journal} {Quantum Sci. Technol.}\ }\textbf {\bibinfo {volume} {4}},\
  \bibinfo {pages} {014002} (\bibinfo {year} {2019})}\BibitemShut {NoStop}%
\bibitem [{\citenamefont {Nakagawa}\ \emph {et~al.}(2020)\citenamefont
  {Nakagawa}, \citenamefont {Tsuji}, \citenamefont {Kawakami},\ and\
  \citenamefont {Ueda}}]{Nakagawa_2020}%
  \BibitemOpen
  \bibfield  {author} {\bibinfo {author} {\bibfnamefont {M.}~\bibnamefont
  {Nakagawa}}, \bibinfo {author} {\bibfnamefont {N.}~\bibnamefont {Tsuji}},
  \bibinfo {author} {\bibfnamefont {N.}~\bibnamefont {Kawakami}},\ and\
  \bibinfo {author} {\bibfnamefont {M.}~\bibnamefont {Ueda}},\ }\bibfield
  {title} {\bibinfo {title} {Dynamical sign reversal of magnetic correlations
  in dissipative hubbard models},\ }\href
  {https://doi.org/10.1103/PhysRevLett.124.147203} {\bibfield  {journal}
  {\bibinfo  {journal} {Phys. Rev. Lett.}\ }\textbf {\bibinfo {volume} {124}},\
  \bibinfo {pages} {147203} (\bibinfo {year} {2020})}\BibitemShut {NoStop}%
\bibitem [{\citenamefont {Tomita}\ \emph {et~al.}(2017)\citenamefont {Tomita},
  \citenamefont {Nakajima}, \citenamefont {Danshita}, \citenamefont {Takasu},\
  and\ \citenamefont {Takahashi}}]{Tomita_2017}%
  \BibitemOpen
  \bibfield  {author} {\bibinfo {author} {\bibfnamefont {T.}~\bibnamefont
  {Tomita}}, \bibinfo {author} {\bibfnamefont {S.}~\bibnamefont {Nakajima}},
  \bibinfo {author} {\bibfnamefont {I.}~\bibnamefont {Danshita}}, \bibinfo
  {author} {\bibfnamefont {Y.}~\bibnamefont {Takasu}},\ and\ \bibinfo {author}
  {\bibfnamefont {Y.}~\bibnamefont {Takahashi}},\ }\bibfield  {title} {\bibinfo
  {title} {Observation of the mott insulator to superfluid crossover of a
  driven-dissipative bose-hubbard system},\ }\href
  {https://doi.org/10.1126/sciadv.1701513} {\bibfield  {journal} {\bibinfo
  {journal} {Sci. Adv.}\ }\textbf {\bibinfo {volume} {3}},\ \bibinfo {pages}
  {e1701513} (\bibinfo {year} {2017})}\BibitemShut {NoStop}%
\bibitem [{\citenamefont {Lange}\ \emph {et~al.}(2018)\citenamefont {Lange},
  \citenamefont {Lenar\u{c}i\u{c}},\ and\ \citenamefont {Rosch}}]{Lange_2018}%
  \BibitemOpen
  \bibfield  {author} {\bibinfo {author} {\bibfnamefont {F.}~\bibnamefont
  {Lange}}, \bibinfo {author} {\bibfnamefont {Z.}~\bibnamefont
  {Lenar\u{c}i\u{c}}},\ and\ \bibinfo {author} {\bibfnamefont {A.}~\bibnamefont
  {Rosch}},\ }\bibfield  {title} {\bibinfo {title} {Time-dependent generalized
  gibbs ensembles in open quantum systems},\ }\href
  {https://doi.org/10.1103/PhysRevB.97.165138} {\bibfield  {journal} {\bibinfo
  {journal} {Phys. Rev. B}\ }\textbf {\bibinfo {volume} {97}},\ \bibinfo
  {pages} {165138} (\bibinfo {year} {2018})}\BibitemShut {NoStop}%
\bibitem [{\citenamefont {Mallayya}\ \emph {et~al.}(2019)\citenamefont
  {Mallayya}, \citenamefont {Rigol},\ and\ \citenamefont
  {De~Roeck}}]{Mallaya_2019}%
  \BibitemOpen
  \bibfield  {author} {\bibinfo {author} {\bibfnamefont {K.}~\bibnamefont
  {Mallayya}}, \bibinfo {author} {\bibfnamefont {M.}~\bibnamefont {Rigol}},\
  and\ \bibinfo {author} {\bibfnamefont {W.}~\bibnamefont {De~Roeck}},\
  }\bibfield  {title} {\bibinfo {title} {Prethermalization and thermalization
  in isolated quantum systems},\ }\href
  {https://doi.org/10.1103/PhysRevX.9.021027} {\bibfield  {journal} {\bibinfo
  {journal} {Phys. Rev. X}\ }\textbf {\bibinfo {volume} {9}},\ \bibinfo {pages}
  {021027} (\bibinfo {year} {2019})}\BibitemShut {NoStop}%
\bibitem [{\citenamefont {Daley}(2014)}]{Daley_2014}%
  \BibitemOpen
  \bibfield  {author} {\bibinfo {author} {\bibfnamefont {A.~J.}\ \bibnamefont
  {Daley}},\ }\bibfield  {title} {\bibinfo {title} {Quantum trajectories and
  open many-body quantum systems},\ }\href
  {https://doi.org/10.1080/00018732.2014.933502} {\bibfield  {journal}
  {\bibinfo  {journal} {Adv. Phys.}\ }\textbf {\bibinfo {volume} {63}},\
  \bibinfo {pages} {77} (\bibinfo {year} {2014})}\BibitemShut {NoStop}%
\bibitem [{\citenamefont {Johansson}\ \emph {et~al.}(2012)\citenamefont
  {Johansson}, \citenamefont {Nation},\ and\ \citenamefont {Nori}}]{Qutip01}%
  \BibitemOpen
  \bibfield  {author} {\bibinfo {author} {\bibfnamefont {J.}~\bibnamefont
  {Johansson}}, \bibinfo {author} {\bibfnamefont {P.}~\bibnamefont {Nation}},\
  and\ \bibinfo {author} {\bibfnamefont {F.}~\bibnamefont {Nori}},\ }\bibfield
  {title} {\bibinfo {title} {Qutip: An open-source python framework for the
  dynamics of open quantum systems},\ }\href
  {https://doi.org/https://doi.org/10.1016/j.cpc.2012.02.021} {\bibfield
  {journal} {\bibinfo  {journal} {Computer Physics Communications}\ }\textbf
  {\bibinfo {volume} {183}},\ \bibinfo {pages} {1760} (\bibinfo {year}
  {2012})}\BibitemShut {NoStop}%
\bibitem [{\citenamefont {Johansson}\ \emph {et~al.}(2013)\citenamefont
  {Johansson}, \citenamefont {Nation},\ and\ \citenamefont {Nori}}]{Qutip02}%
  \BibitemOpen
  \bibfield  {author} {\bibinfo {author} {\bibfnamefont {J.}~\bibnamefont
  {Johansson}}, \bibinfo {author} {\bibfnamefont {P.}~\bibnamefont {Nation}},\
  and\ \bibinfo {author} {\bibfnamefont {F.}~\bibnamefont {Nori}},\ }\bibfield
  {title} {\bibinfo {title} {Qutip 2: A python framework for the dynamics of
  open quantum systems},\ }\href
  {https://doi.org/https://doi.org/10.1016/j.cpc.2012.11.019} {\bibfield
  {journal} {\bibinfo  {journal} {Computer Physics Communications}\ }\textbf
  {\bibinfo {volume} {184}},\ \bibinfo {pages} {1234} (\bibinfo {year}
  {2013})}\BibitemShut {NoStop}%
\bibitem [{\citenamefont {Vitale}\ \emph {et~al.}(2021)\citenamefont {Vitale},
  \citenamefont {Elben}, \citenamefont {Kueng}, \citenamefont {Neven},
  \citenamefont {Carrasco}, \citenamefont {Kraus}, \citenamefont {Zoller},
  \citenamefont {Calabrese}, \citenamefont {Vermersch},\ and\ \citenamefont
  {Dalmonte}}]{vitale2021}%
  \BibitemOpen
  \bibfield  {author} {\bibinfo {author} {\bibfnamefont {V.}~\bibnamefont
  {Vitale}}, \bibinfo {author} {\bibfnamefont {A.}~\bibnamefont {Elben}},
  \bibinfo {author} {\bibfnamefont {R.}~\bibnamefont {Kueng}}, \bibinfo
  {author} {\bibfnamefont {A.}~\bibnamefont {Neven}}, \bibinfo {author}
  {\bibfnamefont {J.}~\bibnamefont {Carrasco}}, \bibinfo {author}
  {\bibfnamefont {B.}~\bibnamefont {Kraus}}, \bibinfo {author} {\bibfnamefont
  {P.}~\bibnamefont {Zoller}}, \bibinfo {author} {\bibfnamefont
  {P.}~\bibnamefont {Calabrese}}, \bibinfo {author} {\bibfnamefont
  {B.}~\bibnamefont {Vermersch}},\ and\ \bibinfo {author} {\bibfnamefont
  {M.}~\bibnamefont {Dalmonte}},\ }\bibfield  {title} {\bibinfo {title}
  {Symmetry-resolved dynamical purification in synthetic quantum matter},\
  }\href@noop {} {\bibfield  {journal} {\bibinfo  {journal} {arXiv:2101.07814}\
  ,\ } (\bibinfo {year} {2021})}\BibitemShut {NoStop}%
\bibitem [{\citenamefont {Apellaniz}\ \emph {et~al.}(2015)\citenamefont
  {Apellaniz}, \citenamefont {Lücke}, \citenamefont {Peise}, \citenamefont
  {Klempt},\ and\ \citenamefont {T{\'{o}}th}}]{Apellaniz2015}%
  \BibitemOpen
  \bibfield  {author} {\bibinfo {author} {\bibfnamefont {I.}~\bibnamefont
  {Apellaniz}}, \bibinfo {author} {\bibfnamefont {B.}~\bibnamefont {Lücke}},
  \bibinfo {author} {\bibfnamefont {J.}~\bibnamefont {Peise}}, \bibinfo
  {author} {\bibfnamefont {C.}~\bibnamefont {Klempt}},\ and\ \bibinfo {author}
  {\bibfnamefont {G.}~\bibnamefont {T{\'{o}}th}},\ }\bibfield  {title}
  {\bibinfo {title} {Detecting metrologically useful entanglement in the
  vicinity of dicke states},\ }\href
  {https://doi.org/10.1088/1367-2630/17/8/083027} {\bibfield  {journal}
  {\bibinfo  {journal} {New Journal of Physics}\ }\textbf {\bibinfo {volume}
  {17}},\ \bibinfo {pages} {083027} (\bibinfo {year} {2015})}\BibitemShut
  {NoStop}%
\end{thebibliography}%

\end{document}